\documentclass[%
 reprint,
 amsmath,amssymb,
 aps,
]{revtex4-2}

\usepackage{float}
\usepackage{slashed}
\usepackage{xcolor}
\usepackage{amsmath}
\usepackage{subcaption}
\usepackage[english]{babel}
\usepackage{lineno, blindtext}
\usepackage{tikz}
\usepackage{adjustbox}
\usepackage{rotating}
\usepackage{rotfloat}
\usepackage[utf8]{inputenc}
\usepackage{graphicx}% Include figure files
\usepackage{dcolumn}% Align table columns on decimal point
\usepackage{bm}% bold math
\newcommand\Tdiag[4]{%
    \multicolumn{1}{|p{#2}|}{\hskip-\tabcolsep
    \begin{tikzpicture}[%
                baseline={(0,-.25\baselineskip)},
                every node/.style={outer sep=0pt,inner sep=#1}]
    \node[minimum width={#2+1\tabcolsep-\pgflinewidth},
        minimum height=2\baselineskip-\pgflinewidth+\extrarowheight,
        use as bounding box] (box) {};
    \draw[line cap=round] (box.north west) -- (box.south east);
    \node[anchor=south west,text width=.75*#2,align=left] at (box.south west) {#3};
    \node[anchor=north east,text width=.75*#2,align=right] at (box.north east) {#4};
\end{tikzpicture}\hskip-\tabcolsep}}

\usepackage{hyperref}
\usepackage{xcolor}
\hypersetup{
  colorlinks   = true, %Colours links
  urlcolor     = red, %Colour for external hyperlinks
  linkcolor    = blue, %Colour of internal links
  citecolor   = blue %Colour of citations
}

\usepackage{threeparttable}
\begin{document}

\preprint{APS/123-QED}

\title{Search for dark matter production in association with the Z$^{\prime}$ boson at the LHC in pp collisions at $\sqrt{s}$ = 8~TeV using Monte Carlo simulations}% Force line breaks with \\
%\thanks{A footnote to the article title}%

\author{S. Elgammal}
 \altaffiliation[sherif.elgammal@bue.edu.eg]{}%Lines break automatically or can be forced with \\
\author{M. A. Louka}%
 \email{Also at Cairo University}
\affiliation{%
 Centre for theoretical physics, The British University in Egypt, Cairo  %\textbackslash\textbackslash
}%

%\collaboration{MUSO Collaboration}%\noaffiliation

\author{A. Y. Ellithi and M. T. Hussein}
\affiliation{%
 Physics Department, Faculty of Science, Cairo University. %\textbackslash\textbackslash
}%

%\collaboration{CLEO Collaboration}%\noaffiliation

\date{\today}% It is always \today, today,
             %  but any date may be explicitly specified

\begin{abstract}
{This analysis presents the possibility for the search for Dark Matter (DM) particles using events with a Z$^{\prime}$ heavy gauge boson and a large missing transverse momentum at the Large Hadron Collider (LHC). 
We consider the muonic decay of Z$^{\prime}$. 
The analyzed Monte Carlo samples were the Open simulated files produced by the CMS collaboration for proton-proton collisions correspond to an integrated luminosity of the LHC run-{\footnotesize I} with 19.7 fb$^{-1}$ at $\sqrt{s} = $ 8 TeV. 
Two scenarios, one simplified benchmark scenario so called Dark Higgs and the effective field theory (EFT) formalism,  were used for interpretations. 
Limits are set on both Z$^{\prime}$, dark matter masses and the cutoff scale of the EFT.}
%This is the first search for DM particles produced in association with a %Z$^{\prime}$ heavy gauge boson decaying to a pair of muons.}

\begin{description}
\item[Keywords]
Dark matter, New heavy gauge boson, The Large Hadron Collider LHC, The Compact Muon Solenoid CMS
\end{description}
\end{abstract}

\maketitle

%\tableofcontents

%=============================================================================

%\begin{acknowledgments}
%We wish to acknowledge the support of the author community in using
%REV\TeX{}, offering suggestions and encouragement, testing new versions,
%\dots.
%\end{acknowledgments}

%\nocite{*}

%================================================

\section{Introduction}
\label{sec:intro}
One of the interesting open questions in modern physics, that can be explored within the current research, is the existence of a new type of non-luminous matter, which can be possibly made up of non-baryonic particles called Dark Matter (DM). 
The need of a DM hypothesis arises from several astrophysical observations and it is supposed that DM contributes to about 27\% of the mass of the Universe \cite{R4, R5, R6, R7, R8, R9, R10, R11, R1010}.
In parallel to the evidence from astrophysical constrains, direct search for DM at the Large Hadron Collider (LHC) is ongoing using proton-proton collision events with a signature based on large missing transverse momentum. 
The methodology of the direct detection of DM at the particle colliders is implemented in the "Mono-X" models, which predict the production of a visible state particle plus missing transverse momentum recoiling against that particle, and is interpreted as the production of dark sector particles with the signature $\text{X}+\slashed{p}_{T}$. 
The visible particle could be a standard model (SM) particle, i.e. W, Z or jets \cite{R35}, photon \cite{photon} or Higgs \cite{R36}. DM particles have also been searched for in events with dilepton from Z boson decay plus large missing transverse momentum with centre-of-mass energies $\sqrt{s}$ = 8 \cite{R450} and 13 TeV \cite{R45055}.\\
The same idea has been extended to beyond the standard model (BSM) particles \cite{R12,R38}.
The model we study (Mono-Z$^{\prime}$) predicts the production of DM in association with the new heavy gauge boson denoted by $Z’$. These dark sector particles can be identified in the detectors located at the LHC as a large missing energy\cite{R1}.
The Z$^{\prime}$ boson is considered as a new particle in many of BSM theories and some extensions of the SM, e.g. the grand unification scenarios. 
The Z$^{\prime}$ is neutral, and has the same decay modes of the standard model Z boson with a larger spectra of masses. More information can be found in \cite{R13,R14,R15,R16}. 
Hence, the search for new heavy resonance in the hadronic and leptonic channels is one of the important goals of high energy physics, in order to investigate those BSM hypotheses that predict new sort of bosons. 
The ATLAS collaboration in \cite{R37} studied the previously mentioned Mono-Z$^{\prime}$ model considering the hadronic decay of Z$^{\prime}$ boson. The coupling of Z$^{\prime}$ to electrons is constrained previously by LEP measurements in \cite{LEP}, thus in our analysis we consider the muonic decay of Z$^{\prime}$ (i.e. Z$^{\prime}~\rightarrow~\mu^{+}\mu^{-}$). 
\\
Thanks to the CMS open data project \cite{R21}, since the CMS collaboration have published significant amounts of recorded and simulated pp collision data at $\sqrt{s} = 8$ TeV, which are available for all scientists even if they are not members of the CMS collaboration. 
In our study we use only the CMS simulated samples. 
These samples have a great potential and offer opportunities to evaluate the cross sections of the SM processes, to model the SM backgrounds for different studies and to perform further analysis, as reported in \cite{R3}.
%as well as the cross sections of some of the SM processes, including the samples used in %this analysis, are re-evaluated in the same reference using some of the open data samples, %and compared to the CMS collaboration published results showing a good consistency between %the two studies.\\
\\
\\
In this paper we first give a more detailed description of the theoretical model which predicts the possible production of DM in association with the heavy gauge boson (Z$^{\prime}$) in section \ref{section:model}, while in section \ref{section:CMS} we briefly describe the Compact Muon Solenoid (CMS) detector.
In section \ref{section:MCandDat} we discuss the simulated samples, for signals and SM background sources, used in the analysis. That is followed in section \ref{section:Backgrounds} by the discussion of the backgrounds in this search and the methods with which their contributions are estimated. 
The selection of events and the general strategy of the search are outlined in section \ref{section:AnSelection}. 
Systematic uncertainties affecting on the prediction of backgrounds are presented in section \ref{section:Uncertainties}. 
The results and summary of the search are respectively addressed in section \ref{section:Results} and section \ref{section:Summary}.

%=================================================
\section{Mono-Z$^{\prime}$ model}
\label{section:model}
\begin{figure} [h]
    \centering
    \subfloat[\centering DH]{{\includegraphics[width=5.5cm]{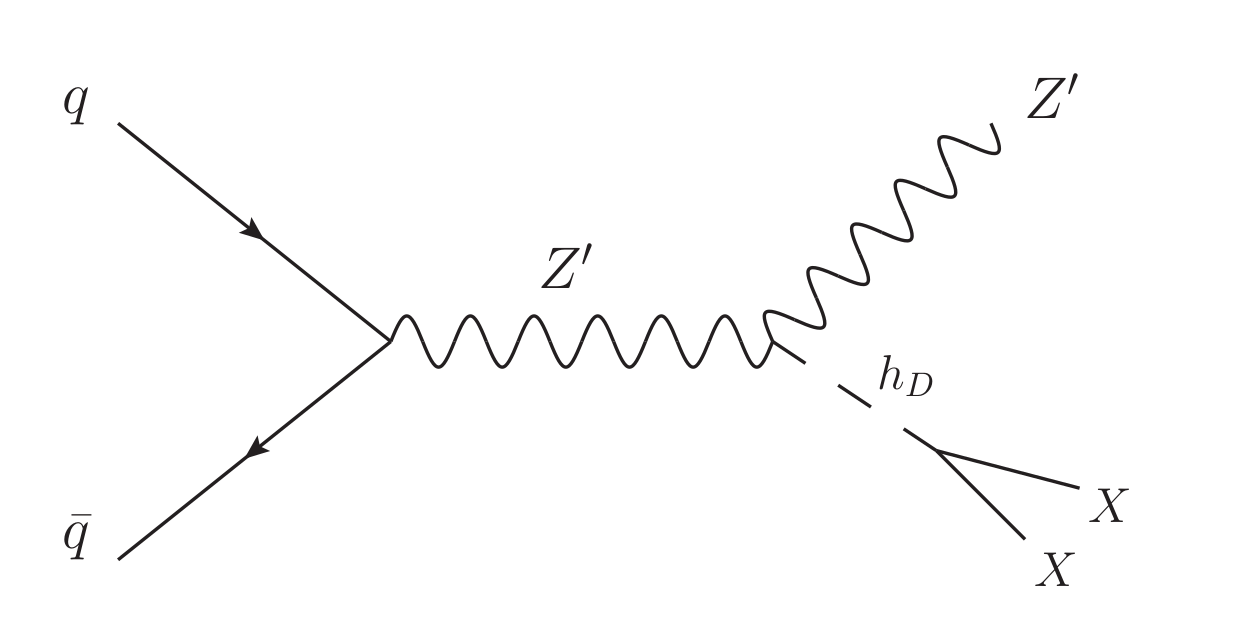} }}%
    \qquad
    \subfloat[\centering LV]{{\includegraphics[width=5.5cm]{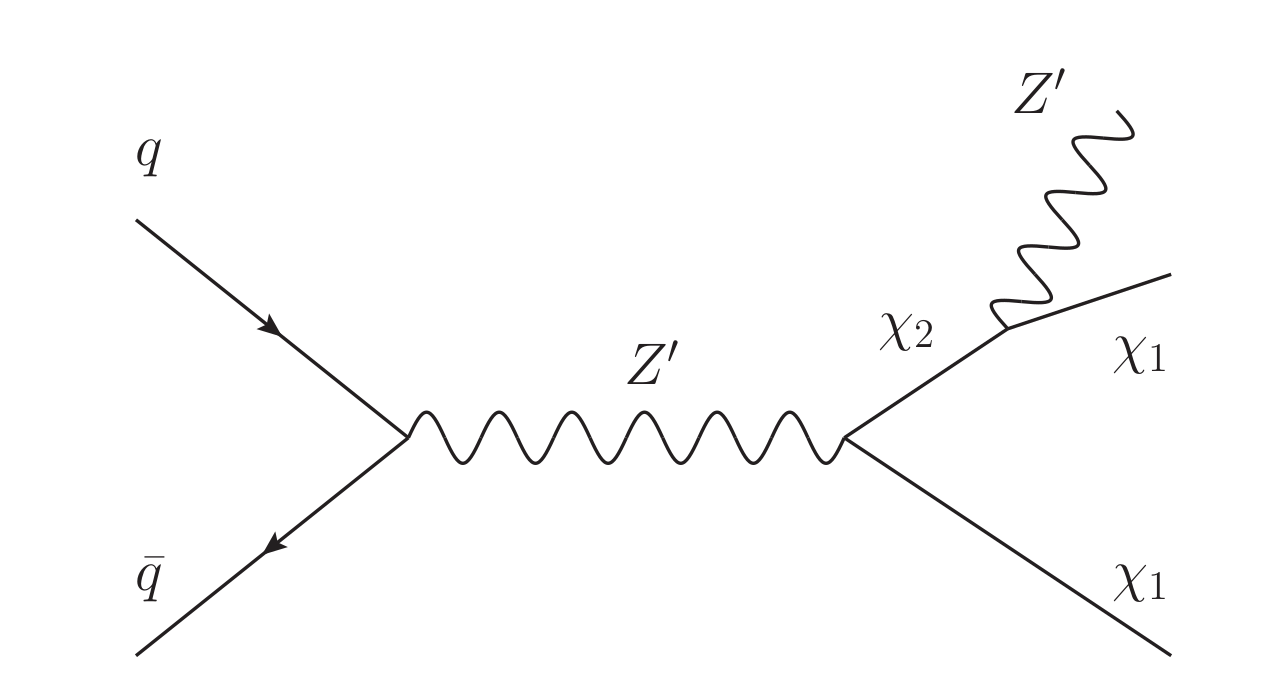} }}%
    \qquad
 \subfloat[\centering EFT]{{\includegraphics[width=5.7cm]{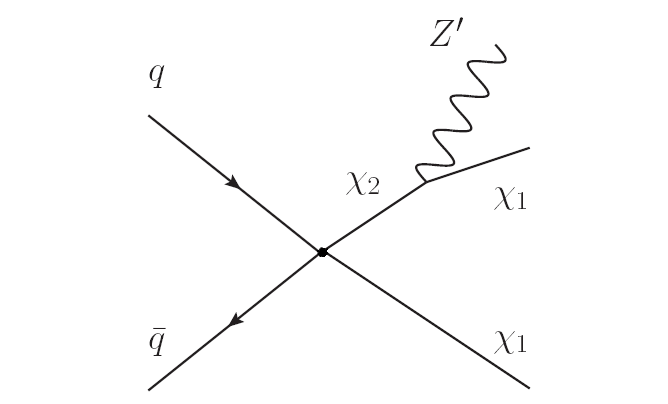} }}%
    \qquad

    \caption{Feynman diagrams for the mono-Z$^{\prime}$ simplified scenarios; dark higgs (a) and light vector (b), and the EFT scenario (c) \cite{R1}.}%
    \label{figure:fig1}%
\end{figure}

The work done in \cite{R1} proposed the production of dark matter with a resonance which comes from heavy Z$^{\prime}$ gauge boson, this model is known as the Mono-Z$^{\prime}$ model. 
The model has been presented in three different possible scenarios which are two simplified models, the dark Higgs (DH) and the light vector (LV) or also called the dark fermion, in addition to a third scenario which is known as light vector with inelastic effective field theory coupling (EFT). The two simplified models are represented in figure \ref{figure:fig1}.
In the DH scenario, the mediator vector boson Z$^{\prime}$ is produced via $q\bar{q}$ annihilation process, at parton level, then it undergoes a dark-higgs-strahlung process analogous to the emission of the SM higgs particle by the W or Z bosons. 
The new scalar coupled to the Z$^{\prime}$ is called the dark higgs ($h_{D}$), and it is assumed that the dark higgs decays rapidly into a pair of dark sector particles ($\chi \bar{\chi}$). 
The coupling of Z$^{\prime}$ with $h_{D}$ is given by (${\fontfamily{qcs}\selectfont{g}_{DM}}M_{{Z}'}h_{D}{Z}'_{\mu}{Z}'^{\mu}$) and the coupling of it with quarks is (${\fontfamily{qcs}\selectfont{g}_{SM}}\bar{q}\gamma^{\mu}q{Z}'^{\mu}$). 
The Feynman diagram of the process is shown in figure \ref{figure:fig1}(a).
% ========  Feynman Diagrams  ==============================
%====================================================
%============================================================
\begin{table} [h!]    %[!hbt]
\centering
\begin {tabular} {ll}
\hline
\hline
Scenario & \hspace{1cm} Masses assumptions \\
\hline

% & \hspace{1cm} $M_{\chi} = 5~\text{GeV}$ \\
  \\
    Light dark sector &  \hspace{1cm} $M_{h_{D}} =$ 
    $\begin{cases}
        M_{Z'}, & M_{Z'} < 125~\text{GeV} \\
        125~\text{GeV}, & M_{Z'} > 125~\text{GeV}
    \end{cases}$ \\
  \\
\hline

% & \hspace{1cm} $M_{\chi}=5$ GeV \\
\\
    Heavy dark sector & \hspace{1cm}  $M_{h_{D}} =$ 
    $\begin{cases}
        125~\text{GeV}, & M_{Z'} < 125~\text{GeV} \\
        M_{Z'}, & M_{h_{D}} > 125~\text{GeV}.
  \end{cases}$\\
 & \\
\hline
\hline
\end {tabular}
\caption{The mass assumptions chosen in the light and heavy dark sector cases for the dark higgs scenario \cite{R1}.}
\label{table:tab1}
\end{table}
%=========================================================
\\
There are two assumptions for setting masses in the DH scenario, which are illustrated for the light dark sector and the heavy dark sector in table \ref{table:tab1}. \\
For the light vector scenario; one of the dark particles is sufficiently heavier than Z$^{\prime}$, so that it can decay to Z$^{\prime}$ plus another light dark particle ($\chi_{2}\rightarrow {Z'}\chi_{1}$) as shown in figure \ref{figure:fig1}(b). The interaction term, in Lagrangian, between the dark particles and Z$^{\prime}$ is given by 
$$ \frac{{\fontfamily{qcs}\selectfont{g}_{DM}}}{2} {Z}'_{\mu}(\bar{\chi}_{2}\gamma^{\mu}\gamma^{5}\chi_{1}~+~\bar{\chi}_{1}\gamma^{\mu}\gamma^{5}\chi_{2}),$$
where $\chi_{1}$ is a final state dark sector stable particle.
For the mass assumptions in case of the LV scenario; the heavy dark particle ($\chi_{2}$) should have a mass twice the mass of Z$^{\prime}$, while the mass of light dark particle ($\chi_{1}$) is half of the mass of Z$^{\prime}$.
\\
In the rest of this paper, the coupling of Z$^{\prime}$ with the SM fermions (quarks and leptons) will be referred to as ${\fontfamily{qcs}\selectfont{g}_{SM}}$, and the coupling with the DM particles will be denoted by ${\fontfamily{qcs}\selectfont{g}_{DM}}$.
The total decay widths of Z$^{\prime}$ and $h_{D}$ (Z$^{\prime}$ and $\chi_{2}$) in the DH (LV) cases, are calculated regarding the mass values of Z$^{\prime}$ and the coupling constants, assuming that Z$^{\prime}$ boson can only decay into a pair of muons and radiate an $h_{D}$ boson in the DH scenario, and assuming that the decays ${Z}'\rightarrow\chi_{1}\chi_{2}$, $\chi_{2}\rightarrow {Z'}\chi_{1}$ and ${Z}'\rightarrow\mu\tilde{\mu}$ are the only allowed for the LV scenario.
\\
In these scenarios there are many free parameters, including the mediator mass $M_{Z^{\prime}}$, the mass of the heavy dark particle $M_{\chi_{2}}$, the mass of the light dark particle $M_{\chi_{1}}$ and the coupling constants (${\fontfamily{qcs}\selectfont{g}_{SM}}$ and ${\fontfamily{qcs}\selectfont{g}_{DM}}$). 
In this analysis, the values of the couplings (${\fontfamily{qcs}\selectfont{g}_{SM}} = 0.25$ and ${\fontfamily{qcs}\selectfont{g}_{DM}} = 1.0$) have been chosen based on the results presented in \cite{R1} and \cite{R37}. 
The cross  section measurements times branching ratios for the two simplified models (DH and LV) at different masses of Z$^{\prime}$ are compared in table \ref{table:simplfied}, and are calculated with Madgraph \cite{R33} at next-to-leading order (NLO).\\
Finally, the EFT scenario reduces the interactions between the DM particles and the SM fields down to contact interaction as given in the following interaction term
$$ \frac{1}{2\Lambda^{2}}\bar{q}\gamma^{\mu}q(\bar{\chi}_{2}\gamma^{\mu}\gamma^{5}\chi_{1}~+~\bar{\chi}_{1}\gamma^{\mu}\gamma^{5}\chi_{2})$$
The Feynman diagram, which illustrates this process, is shown in figure \ref{figure:fig1}(c). The assumptions for the masses are the same set of the LV scenario. The production cross section measurements times branching ratios as a function of the scenario cutoff scale ($\Lambda$) are given in table \ref{table:EFT}.
% ================================================
% ============ cross section tables  ===============================
%=========================================
%= Sigma Vs. Mzp 
%=========================================
\begin{table}[h!]
\centering
%\label{ tab-marks }
%\begin{subt}
\begin {tabular} {|c|c|c|}
\hline
$M_{{Z}'}$ & $\sigma~\times$ BR~~(pb) & $\sigma~\times$ BR~~(pb) \\
\big(GeV/c$^{2}$\big)             & Dark Higgs & Light Vector  \\
\hline
\hline
150 & $7.086~~ \times 10^{-2}$ & $1.734~~ \times 10^{-2}$  \\
\hline
200 & $2.366~~ \times 10^{-2}$ & $0.507~~\times 10^{-2}$ \\
\hline
250 & $9.555~~ \times 10^{-3}$ & $1.808~~ \times 10^{-3}$ \\
\hline
300 & $4.368~~ \times 10^{-3}$ & $0.738~~\times 10^{-3}$  \\
\hline
350 & $2.100~~ \times 10^{-3}$ &  $0.318~~\times 10^{-3}$  \\
\hline
400 & $1.040~~ \times 10^{-3}$ & $0.140~~\times 10^{-3}$  \\
\hline
450 & $0.569~~ \times 10^{-3}$ & $0.069~~ \times 10^{-3}$   \\
\hline
500 & $3.283~~ \times 10^{-4}$ & $0.355~~ \times 10^{-4}$   \\
\hline
600 & $1.191~~ \times 10^{-4}$ & $0.104~~ \times 10^{-4}$   \\
\hline
700 & $4.725~~ \times 10^{-5}$ &  $0.333~~ \times 10^{-5}$ \\
\hline
\end {tabular}
\caption{The cross section measurements times branching ratios calculated at different masses of the Z$^{\prime}$ boson with the heavy dark sector assumption for the two simplified models (DH and LV), with the coupling constants ${\fontfamily{qcs}\selectfont{g}_{SM}} = 0.25,~{\fontfamily{qcs}\selectfont{g}_{DM}} = 1.0$, at $\sqrt{s} = 8$ TeV.}
\label{table:simplfied}
\end{table}
%=======================================================
\begin{table} [h!]
\centering
\begin{tabular}{|c|c|}
\hline
    $\Lambda (TeV)$ & $\sigma~\times$ BR~~(pb)  \\
\hline
\hline
    1.0 & 0.0704  \\
\hline
    1.5 & 0.0139 \\
\hline
    2.0 & 0.0044  \\
\hline
    2.5 & 0.0018 \\
\hline
     3.0& 0.00087  \\
\hline
     3.5& 0.00047  \\
\hline
     4.0& 0.000275  \\
\hline
\end{tabular}
\caption{The EFT production cross section measurements times branching ratios as a function of the scenario cutoff scale of the EFT($\Lambda$), for a fixed mass point of Z$^{\prime}$ ($M_{Z^{\prime}}$ = 450 GeV) and centre-of-mass energy $\sqrt{s} = 8$ TeV.}
\label{table:EFT}
\end{table}

\begin{figure} [h]
\centering
\resizebox*{9cm}{!}{\includegraphics{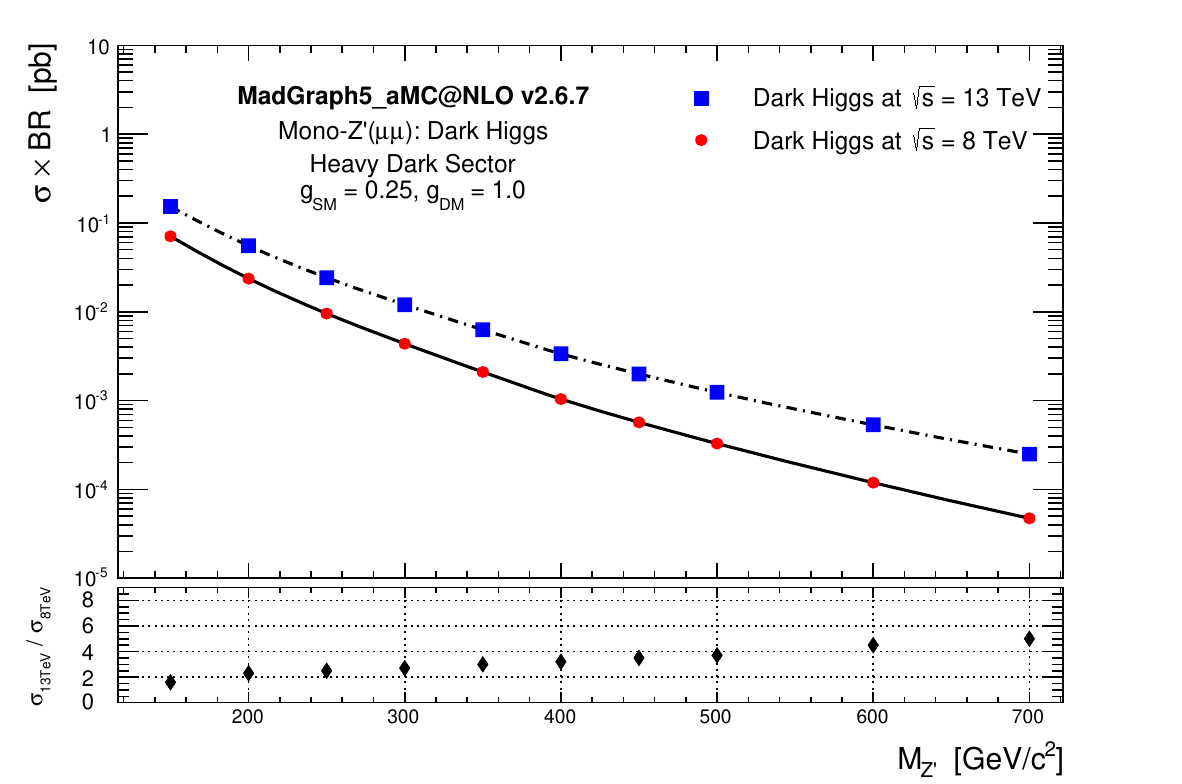}}
\caption{The behavior of the cross section measurements times branching ratios with the mass of $Z'$ boson for the Dark Higgs scenario at $\sqrt{s} = 8$ TeV represented by red dots and 13 TeV represented by blue square.
The lower panel indicates to the ratio of the cross sections between the two cases.}
\label{figure:fig2}
\end{figure}

%=======================================================
The typical signature of these processes consists of a pair of opposite sign leptons or hadronic jets from the decay of Z$^{\prime}$ plus a large missing transverse momentum due to the stable dark sector particles $\chi$ and $\chi_{1}$. 
These two scenarios were previously studied by the ATLAS collaboration in \cite{R37} with the hadronic decay of Z$^{\prime}$. 
In our study, we have considered the heavy dark sector assumption mentioned in table \ref{table:tab1}, at which the signal region is shifted away from the background region due to a larger missing energy that assumed in this option, i.e the signal is more distinguishable from the background for the heavy dark sector assumption rather than the light one. 
The muonic decay of the on-shell Z$^{\prime}$ is considered since the CMS detector has been optimized to this decay channel. 
So that our the events are with the following topology $ \mu^{+}\mu^{-} +\slashed{E}_{T}$.\\
We studied one of these two simplified models which is the DH scenario, since it has a higher cross section than the LV, in addition to the EFT scenario.
The behavior of the cross sections times branching ratios with the mass of Z$^{\prime}$, at $\sqrt{s}$ = 8 and 13 TeV for the DH scenario, is shown in figure \ref{figure:fig2}.
As expected, the cross section measurements times branching ratios drop with the increase of the Z$^{\prime}$ mass. Moreover, we observe the increase in the cross section at higher $\sqrt{s}$ (13 TeV), which is an advantage of the LHC run-{\small II} data with respect to run-{\small I}, as well as the ratio between the cross sections in the two cases increases with the mass of Z$^{\prime}$, and reaches its maximum value (around 5 times) for the scanned range at $M_{Z'}= 700$ GeV, which indicates that the cross section drops slower in the case of $\sqrt{s}=$13 TeV. \\
Table \ref{table:tabchi} indicates the cross section measurements times branching ratios calculated for different sets of the Z$^{\prime}$ and $\chi$ masses. The cross section is not sensitive to the change in the dark matter particle mass, for this reason we work on the diagonal points for our purpose to put a limit on this parameter which will be discussed in the results. 
% ================ figure ==========================
%===============================================================
%========  sigma Vs Mchi ===========================
\begin{table} [p]
%\rotatebox{90}{
\begin{sidewaystable} [H]
%\begin{adjustbox}{angle=90}
%\begin{ruledtabular}
\centering
%\begin{flushleft}
%\tiny
%\scriptsize
%\fontsize{3pt}{4pt}
%\selectfont
\begin{tabular}{|c||c|c|c|c|c|c|c|c|c|c|c|}
\hline
%\backslashbox{$M_{\chi}$ (GeV)}{$M_{Z'}$ (GeV)}
\Tdiag{.01em}{1.6cm}{$M_{\chi}$\tiny{(GeV)}}{$M_{Z'}$\tiny{(GeV)}}  & 150 & 200 & 300 & 325 & 350 & 375 & 400 & 425 & 450 & 475 & 500 \\
\hline
\hline
1  & $7.10\times10^{-02}$ & $2.36\times10^{-02}$ & $0.438\times10^{-02}$ & $0.305\times10^{-02}$  & $0.2107\times10^{-02}$ & $0.144\times10^{-02}$ & $0.1036\times10^{-02}$ & $0.764\times10^{-03}$ & $0.568\times10^{-03}$ & $0.428\times10^{-03}$ & $0.328\times10^{-03}$ \\
\hline
5  & $7.08\times10^{-02}$ & $2.37\times10^{-02}$ & $0.437\times10^{-02}$ & $0.306\times10^{-02}$ & $0.210\times10^{-02}$ & $0.144\times10^{-02}$ & $0.104\times10^{-02}$ & $0.763\times10^{-03}$ & $0.569\times10^{-03}$ & $0.427\times10^{-03}$ & $0.3283\times10^{-03}$\\
\hline
10 & $7.10\times10^{-02}$ & $2.36\times10^{-02}$ & $0.037\times10^{-02}$ & $0.305\times10^{-02}$ & $0.211\times10^{-02}$ & $0.145\times10^{-02}$ & $0.104\times10^{-02}$ & $0.763\times10^{-03}$ & $0.569\times10^{-03}$ & $0.429\times10^{-03}$ & $0.328\times10^{-03}$\\
\hline
25 & $7.10\times10^{-02}$ & $2.358\times10^{-02}$ & $0.437\times10^{-02}$ & $0.305\times10^{-02}$ & $0.211\times10^{-02}$ & $0.144\times10^{-02}$ & $0.1035\times10^{-02}$ & $0.763\times10^{-03}$ & $0.568\times10^{-03}$ & $0.429\times10^{-03}$ & $0.329\times10^{-03}$\\
\hline
50 & $7.13\times10^{-02}$ & $2.36\times10^{-02}$ & $0.437\times10^{-02}$ & $0.306\times10^{-02}$  & $0.210\times10^{-02}$ & $0.144\times10^{-02}$ & $0.1038\times10^{-02}$ & $0.756\times10^{-03}$  & $0.567\times10^{-03}$ & $0.429\times10^{-03}$ & $0.328\times10^{-03}$\\
\hline
75 & $16.40\times10^{-02}$  & $2.36\times10^{-02}$ & $0.436\times10^{-02}$ & $0.305\times10^{-02}$ & $0.209\times10^{-02}$ & $0.144\times10^{-02}$ & $0.104\times10^{-02}$  & $0.763\times10^{-03}$ & $0.568\times10^{-03}$ & $0.429\times10^{-03}$ & $0.328\times10^{-03}$ \\
\hline
100 & $8.98 \times10^{-07}$ & $5.43\times10^{-02}$ & $0.436\times10^{-02}$ & $0.3052\times10^{-02}$ & $0.211\times10^{-02}$ & $0.144\times10^{-02}$ & $0.1039\times10^{-02}$ & $0.764\times10^{-03}$ & $0.568\times10^{-03}$ & $0.428\times10^{-03}$ & $0.326\times10^{-03}$ \\
\hline
125 & $1.54\times10^{-07}$ & $5.01\times 10^{-07}$ & $0.437\times10^{-02}$ & $0.3049\times10^{-02}$ & $0.209\times10^{-02}$ & $0.144\times10^{-02}$ & $0.104\times10^{-02}$ & $0.758\times10^{-02}$ & $0.567\times10^{-03}$ & $0.429\times10^{-03}$ & $0.327\times10^{-03}$ \\
\hline 
150 & $4.09 \times10^{-08}$ & $1.0 \times10^{-07}$ & $1.00\times10^{-02}$ & $0.3047\times10^{-02}$ & $0.2094\times10^{-02}$ & $0.144\times10^{-02}$ & $0.1037\times10^{-02}$ & $0.758\times10^{-03}$ & $0.567\times10^{-03}$ & $0.427\times10^{-03}$ & $0.326\times10^{-03}$ \\
\hline
175 & $1.3\times10^{-08}$ & $3.05\times10^{-08}$ & $1.753 \times10^{-07}$ & $3.766 \times10^{-07}$ & $0.48\times10^{-02}$ & $0.1436\times10^{-02}$ & $0.1035\times10^{-02}$ & $0.756\times10^{-03}$ & $0.556\times10^{-03}$ & $0.427\times10^{-03}$ & $0.326\times10^{-03}$ \\
\hline
200 & $5.25 \times10^{-09}$ & $1.0 \times10^{-08}$ & $4.36 \times10^{-08}$ & $6.56 \times10^{-08}$ & $1.04 \times10^{-07}$ & $2.08 \times10^{-07}$ & $0.239\times10^{-02}$ & $0.76\times10^{-03}$ & $0.566\times10^{-03}$ & $0.427\times10^{-03}$ & $0.325\times10^{-03}$ \\
\hline
\end {tabular}

\caption{ The Dark Higgs cross section measurements times branching ratios (pb) calculated for different sets of the masses $M_{\chi}$ and $M_{Z'}$ in GeV, for the heavy dark sector mass assumption, with the following couplings constants $g_{SM} = 0.25,~g_{DM} = 1.0$ and $\sqrt{s} = 8$ TeV.}
\label{table:tabchi}
\end{sidewaystable}
%\end{flushleft}
%\end{ruledtabular}
%\end{adjustbox}
\end{table}
%\end{sidewaystable}

%=========================================================
%%%%\newpage
%$~~$
%\newpage
\section{The CMS detector}
\label{section:CMS}
The Compact Muon Solenoid (CMS) detector (described in details at references \cite{R17,R29}) is one of the four main apparata that have been built on the LHC at CERN. The 3-m-long and 5.9-m-inner diameter superconducting conducting solenoid provides a 3.8-T magnetic field that delivers the bending power required to measure the momenta of the high energy charged particles.
The solenoid accommodates the tracking system (pixel detector and silicon tracker) and the two calorimeters: the Electromagnetic Calorimeter (ECAL) has been designed to detect and measure electrons and photons and the Hadronic Calorimeter (HCAL) used to detect and measure the hadronic particles. 
The muon system envelope the above layers, the muon stations consist of many Drift Tubes (DT) layers in the barrel part and Cathode Strip Champers (CSC) in the endcape region. 
The two parts are completed by the Resistive Plate Champers (RPC).\\
The interaction point is considered to be the origin of the CMS coordinate system. The \text{x}-axis points toward the center of the LHC, the \text{y}-axis points upward and the \text{z}-axis is alongside the beam axis.
The polar angle $\theta$ is measured from the positive direction of the \text{x}-axis, and the azimuthal angle $\phi$ is measured from the \text{x-y} transverse plan. However, the directions of the particles yield from the collision spot mostly expresses in terms of the pseudorapidity which defined as 
$ \eta = - \text{ln}[\text{tan}(\theta/2)] $.
For our purpose, we mention the reconstruction of muons and the missing transverse momentum. The muon object is identified and reconstructed from a global fit between the muon system and the inner tracker hence, it is referred to as global muons \cite{R18,R40}.
The missing transverse momentum reconstruction is based on the Particle Flow algorithm described in references \cite{R19,R29}, 
it is reconstructed as an imbalance in the vector sum of momenta in the transverse plan. i.e. it could be defined as the negative vector sum of the momenta of all particle flow reconstructed objects as 
$ \vec{\slashed{p}}_{T} = -\sum \vec{p}_{T}^{~pf} $ \cite{R45}.
The magnitude of $\vec{\slashed{p}}_{T}$ can be affected by many factors which cause underestimate or overestimate of its true value. 
These factors are basically related to the calorimeters response, 
as minimum energy and $p_{T}$ thresholds in the calorimeters, and inefficiencies in the tracker, and non-linearity of the response of
the calorimeter for hadronic particles.
This bias can be effectively reduced by correcting for the $p_{T}$
of the jets using jet energy corrections as defined in the following formula, which is given in \cite{R45} 
$$ 
\vec{\slashed{p}}_{T}^{~\text{corr}} = \vec{\slashed{p}}_{T} -
\sum_{jets} (\vec{p}_{T jet}^{~\text{corr}} - \vec{p}_{T jet}),
$$
where "corr" refers to the corrected values. 
Thus variables of particular relevance to the present analysis are the corrected missing transverse momentum vector $\vec{\slashed{p}}_{T}^{~\text{corr}}$ and the magnitude of this quantity, $\slashed{\slashed{p}}_{T}^{~\text{corr}}$.

%As long as the muon momentum is much higher than its mass, one can %consider that its transverse momentum and transverse energy are equal 

%===========================================================
\section{Simulated samples}
\label{section:MCandDat}
\subsection{Monte Carlo simulation of the model signals}
The model signal events are generated using \text{MadGraph5\_aMC@NLO}~v2.6.7 \cite{R33} which is a general purpose matrix element event generator. The cross section calculated at next to-leading-order (NLO) and the hadronizaton process has been done with \text{Pythia} \cite{R34}. The detector simulation, simulation of read out system response (digitization) and reconstruction processes have been done using the standard CMS open data software framework (the release \text{CMSSW\_5\_3\_32}) at $\sqrt{s} = $ 8 TeV requirements, with the suitable triggers list used for CMS-2012 analysis.
We scanned the DH production cross section at different sets of the masses of the particles Z$^{\prime}$ and ${\chi}$ as free parameters covering a wide range for the mass of Z$^{\prime}$ boson from 150 GeV to 550 GeV and from 1 GeV to 200 GeV for the mass of ${\chi}$, and the production cross section of the EFT at the range of $\Lambda$ from 1.0 to 4 TeV, assuming $\fontfamily{qcs}\selectfont{g}_{SM} = 0.25$ and $\fontfamily{qcs}\selectfont{g}_{DM} = 1.0$ for all simulations.
%The calculations of the cross section times branching ratio, for the DH and LV with different masses of Z$^{\prime}$, and for EFT scenarios at {$\sqrt{s}$} equal to 8 TeV are indicated in table \ref{table:tab2}. 

%\begin{table}[h!]
%\centering
%\begin{tabular}{|c||c|c|c|c|c|c|c|c|c|c|c|}\hline
%\hline
%$M_{chi}$ (GeV) & 1 & 5 & 10 & 25 & 50 & 75 & 100 & 125 & 150 & 175 & 200 \\
%$M_{Z'}$ (GeV) & 150 & 200 & 300 & 325 & 350 & 375 & 400 & 425 & 450 & 475 & 500 \\
%\hline
%\hline
%$\sigma~\times$ BR~~(pb) & & & & & & & & & & & \\
%\hline
%\end {tabular}
%\caption{ The Dark Higgs cross section measurements times branching ratio calculated for different masses of the dark particle $\chi$, for the heavy dark sector mass assumption, with the following couplings constants $g_{SM} = 0.25,~g_{DM} = 1.0$}
%\label{table:tabchi2}
%\end{table}

%======================================================
%%%%\newpage
\subsection{Monte Carlo simulation of the SM backgrounds}
In order to simulate the SM processes that have muons and/or missing energy (due to the undetected neutrinos) at the final state which could interface with our signal events, we used the CMS open Monte Carlo samples at $\sqrt{s}$ = 8 TeV as background processes \cite{R21}.
The Drell-Yan (DY) background (the production of a virtual $Z/\gamma^{*}$ that decay into a muon pair) which is the dominant background, has been generated using \text{Powheg} \cite{powheg}.
Another important sources of SM background with dimuons in the final state are the fully leptonic decay of $t\bar{t}$ which is generated using \text{MadGraph} \cite{R33}, the production of electroweak diboson channels as WW, WZ have been generated with \text{MadGraph}, and $ZZ \rightarrow \mu^{-}\mu^{+}\mu^{-}\mu^{+}$ process which is also generated with \text{Powheg}.
The generation process for the mentioned samples are interfaced with \text{Pythia v6.4.26} \cite{R34} for the modeling of parton shower.
The Monte Carlo samples used in this analysis and their corresponding cross sections, calculated at next-to-leading or next-to-next-to-leading order, are indicated in table \ref{table:tab3}. 

%The cross sections order of calculation (at next-to-leading or %next-to-next-to-leading order) is also stated.
%===========================================================

%======= MCS data Sets ===================================
\begin{table*}[]
\centering
\begin{ruledtabular}
\begin {tabular} {|l|l|l|c|l|}
Process \hspace{0.5cm} & Generator \hspace{0.5cm}  & Data Set Name & ${\Large \sigma} \times~\text{BR}$ ~(\text{pb}) & Order \\
\hline
\hline
$DY \rightarrow \mu\bar{\mu}$  & Powheg & DYToMuMu\_M-20\_CT10\_TuneZ2star\_v2\_8TeV. \cite{R22} \hspace{0.5cm}  & 1916 & NNLO\hspace{0.5cm}\\
\hline
$t\bar{t}$ + jets & Madgraph  & TTJets\_FullLeptMGDecays\_8TeV. \cite{R23} & 23.89 & NLO \\
\hline
%tw & Powheg & 11.1 \\
%\hline
%$\tilde{t}w$ & Powheg & 11.1 \\
%\hline
WW + jets & Madgraph & WWJetsTo2L2Nu\_TuneZ2star\_8TeV. \cite{R24} & 5.8 & NLO \\
\hline
WZ + jets & Madgraph  & WZJetsTo3LNu\_8TeV\_TuneZ2Star. \cite{R25} &1.1 & NNLO \\
\hline
$ZZ\rightarrow 4\mu$  & Powheg & ZZTo4mu\_8TeV. \cite{R26} & 0.077 & NLO \\
\end {tabular}
%\vspace{3}
\caption{CMS open MC samples used to simulate the SM background for pp collision at $\sqrt{s} = 8$ TeV, their corresponding cross section times branching ratio for each process and the order of calculations. The data set names and the used generators are stated.}
\label{table:tab3}
\end{ruledtabular}
\end{table*}
%=====================================================
%=========================================================
%%%%\newpage
\section{Backgrounds estimation}
\label{section:Backgrounds}
There are three main types of SM background to new physics search in the dimuon channel.
The most significant is the irreducible SM Drell-Yan process. New physics can interfere with this process, if not mitigated, the effects can be significant.
The second most important background type comes from muons from non-singularly produced W and Z bosons. 
The dominant source of these muons are from $t\bar{t}$ events although WW events become increasingly important at high mass as the boost of the top quark means that the b-jet enters the muons isolation cone and so the muon fails isolation requirements.
Other sources include WZ and ZZ events although they are small compared to $t\bar{t}$ and WW.
%, mainly entering at the Z peak. 
This background is referred to as $t\bar{t}$ and $t\bar{t}$-like background as it is dominated by $t\bar{t}$.
The third background is the jets background, where one or more jets are misidentified and incorrectly reconstructed as a muon, mainly arising from W+jet and QCD multijet. 
The contamination of jets background is usually estimated from data using a so called data driven method which is explained in \cite{zprime}, which is irrelevant for our study, since our analysis based mainly on MC simulations.  
%It has been found that the QCD mltijet and W+jet contributions are very small %above 400 GeV at the dimuon invariant mass spectrum, as estimated in %\cite{zprime}, with only 3 events are misidentified as muons for an integrated %luminosity of 20.6 fb$^{-1}$, hence in our case (luminosity = 19.7 fb$^{-1}$) %this contribution is expected to be lower than 3 events and is negligible in the %current study.
In this analysis; the DY background and other backgrounds from muons arising from non-singularly produced W and Z bosons is estimated directly from Monte Carlo simulation as in the other 8 TeV similar analysis \cite{zprime}. 
The leptonic decay of $t\bar{t}$, WW and WZ are generated using Madgraph, while ZZ is generated using POWHEG. 
The Monte Carlo samples and their cross sections are documented in tables \ref{table:tab3}, and normalized to their respective NLO or NNLO cross sections. 
Another source of background to this analysis is the cosmic muons contribution, which is suppressed by constraining the vertex position as discussed in section \ref{section:Preliminary}, and also by constraining the impact parameter of the muons relative to the vertex position as in the high $P_{T}$ muon identification \cite{R41, R32}, this contribution is also negligible.
%==============================================================
\section{Events selection and analysis strategy}
\label{section:AnSelection}
\subsection{Preselection of events}
\label{section:Preliminary}
%Preselection criterion is base on the following set of cuts; 
The preselection criteria is based on the high $p_{T}$ muon identification \cite{R41, R32}, which was applied in the 2012 analysis for the search for heavy resonances in the dilepton channel \cite{zprime}; in addition the off-line muon reconstructed transverse momentum $(p^{\mu}_{T})$ is selected to be higher than 45 GeV in order to be fully efficient for the trigger used (\text{HLT\_Mu40\_eta2p1}), and the detector acceptance 
is restricted to the range $|\eta^{\mu}| < 2.1$ of the reconstructed pseudorapidity.
The preselection criteria is indicated in table \ref{table:tabID1}, at which muon candidates must be reconstructed as "global" muons, i.e. standalone muon object reconstructed in the muon system must match with an inner tracker’s track to form the global muon object used later for our analysis.
The muon candidates should be isolated; thus they have to pass a cut based on the
relative tracker isolation, which is the scalar sum of the $p_{T}$ of all other tracks within a cone of $\Delta R = \sqrt{(\Delta\eta)^2 + (\Delta\phi)^2} < 0.3$ around and not containing the muon’s tracker track, this sum must be less than $10\%$ of the muon’s transverse momentum ($p_{T}$).
%, also as measured by the tracker. 
Tracks used in the tracker isolation calculation have to originate within 
$\Delta Z = 0.2$ cm of the primary vertex, with which the muon candidate is associated \cite{zprime}.

\begin{figure}[h]
\centering
\resizebox*{9cm}{!}{\includegraphics{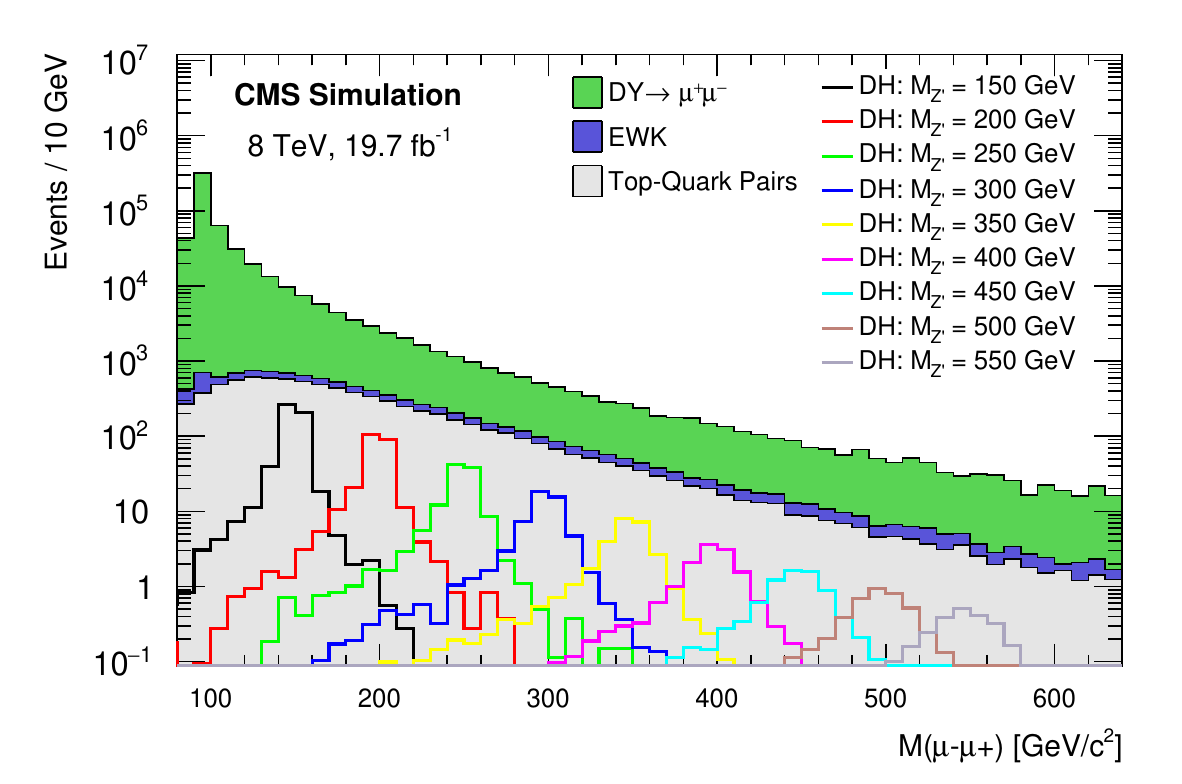}}
\caption{The measured dimuon invariant mass spectrum, after applying preselection cuts listed in table \ref{table:tabID1}, together with the estimated SM backgrounds and Z$^{\prime}$ masses produced according to the dark Higgs model.} 
%The total systematic uncertainty in the overall background is shown as a shaded region. The data-to-simulation ratio is shown in the lower panel.}
\label{figure:fig3}
\end{figure}

\begin{figure}[h]
\centering
\resizebox*{9cm}{!}{\includegraphics{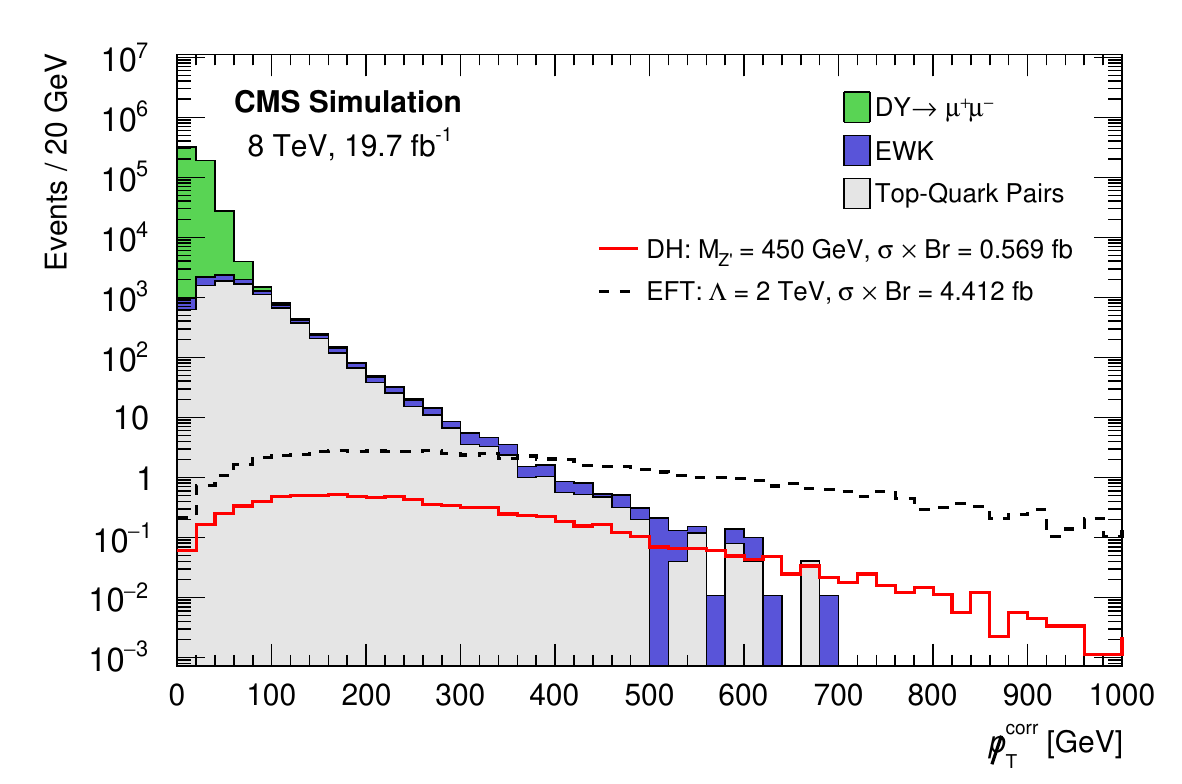}}
%\resizebox*{9cm}{!}{\includegraphics{ptmiss.pdf}}
  %\caption{The distribution of the missing transverse momentum after %the application of the preliminary event selection for the dark %higgs model (heavy dark sector assumption) at $M_{Z}’= 450~GeV$.}
\caption{The missing transverse momentum distribution, after the preselection cuts listed in table \ref{table:tabID1}; the colored stacked histograms refer to the MC simulation of the SM backgrounds, with two signals presentation of the model corresponding to the DH scenario with $M_{Z^{\prime}}=450$ GeV and to the EFT scenario with $\Lambda=2$ TeV are superimposed. The signals are normalized to the product of cross section times the Z$^{\prime} \rightarrow \mu^{+}\mu^{-}$ branching ratio.}
%The lower panel illustrates the ratio between the data and the estimated SM backgrounds.}
\label{figure:fig4}
\end{figure}

%Track isolation cut is used as the following: the transverse momenta scalar sum for tracks %inside a cone of $\Delta R = \sqrt{(\Delta\eta)^2 + (\Delta\phi)^2} < 0.3 $ around the %muon’s track in the inner tracker must be less than $10\%$ of the muon’s transverse %momentum, and the tracks which included in this sum must be within $\Delta Z = 0.2$ 
%cm from the primary vertex associated with that muon.
%The primary vertex must be within the radial distance $|r| < 0.2$ cm from the nominal %interaction point, which is a powerful cut to reject the cosmic muons that pass at the %empty time between two bunch-crossing.
The muon’s transverse impact parameter with respect to the primary vertex, as measured by the tracker-only fit, must be smaller than 0.2 cm, which is a powerful cut to reject the cosmic muons that pass at the empty time between two bunch-crossing.
Another cut is provided to reject cosmic muons that pass near the interaction point in-time with a bunch-crossing, is the 3D angle between the muon pairs, it is selected to be less than $\pi - 0.02$ rad \cite{zprime}.
Extra qualification cuts are applied such that, the muon pairs must be two opposite-sign and the $\chi^{2}/\text{dof}$ for the common vertex fitting is less than 10 \cite{zprime}, where $\chi$ is a state vector that describe the particle's track in each point of its trajectory, this method based on the minimization of $\chi^{2}/\text{dof}$ using the Kalman filter technique, described in references \cite{R30,R31}, is implemented in the CMSSW, this fitting is important in order to make correct pairing of muons that originate from the same vertex and for the rejection of the pile-up muons.
Thus the events are selected with two opposite charge high $p_{T}$ muons, with one of them passed the single muon trigger HLT$\_$Mu40$\_$eta2p1. 
%Each event is weighted by a scale factor (SF) corresponding to efficiency %(including triggering, reconstruction, and identification) ratios between data %and the simulation, this SF was found to be $0.990 \pm 0.005$ in the barrel part %and $0.993 \pm 0.005$ in the endcaps part of the CMS muon detector %\cite{zprime}.
%, of the  respectively.
%As it has been mentioned in section \ref{section:CMSopenData}; the efficiency of %this single muon trigger varies from 97\% to 100\% as function of $\eta^{\mu}$, %the effect of trigger efficiencies on simulated event samples is included by %using the trigger efficiencies determined from the data and applying a weight to %each simulated event. 
%==== Preliminary selection ================
\begin{table}[h]
    \centering
    %\begin{tabular}{|ll|}
    \begin {tabular} {|c|c|}
\hline
 variable & cut value \\
\hline
    \hline
     %Trigger     & HLT\_Mu40\_eta2p1 \\
     High $p_{T}$ muon ID & \cite{R41, R32}\\
     $p^{\mu}_{T}$ (GeV) & $>$ 45 \\
     $|\eta^{\mu}|$ (rad) & $<$ 2.1 \\
     $M_{\mu^{+}\mu^{-}}$ (GeV) & $>$ 50 \\

    \hline
    \end{tabular}
    \caption{The preselection selection criteria based on single muon 
    trigger requirement \text{(HLT\_Mu40\_eta2p1)}, muon kinematic cuts 
    and the high $P_{T}$ muon ID.}
    \label{table:tabID1}
\end{table}
\\
Figure \ref{figure:fig3} illustrates the distribution of the dimuon invariant mass; the green histogram represents the Drell-Yan background, the blue histogram stands for the vector boson pair backgrounds (WW, WZ and ZZ) and the $t\bar{t}$ + jets background is represented by the grey histogram. 
These background histograms are stacked, while the signal models with different masses of the Z$^{\prime}$ heavy boson are represented by different colored lines, and are overlaid. 
%The total systematic uncertainty (explained in section %\ref{section:Uncertainties}) is illustrated in the ratio plot.
The corresponding distribution of the missing transverse momentum is shown in figure \ref{figure:fig4}. 
%These figures show good matching between the data points and the simulated SM backgrounds within the statistical error (demonstrated by the error bars on the data points) and within the systematic uncertainty (demonstrated by the hatched region in the ratio plots). 
As the signal models are overwhelmed by the backgrounds, it is necessary to apply a more clever set of cuts to discriminate signals from SM backgrounds as will be explained in the next section.\\ 
The number of dimuon events passing the preselection for each SM background processes and for the model signals are quoted in table \ref{table:tab6} for an integrated luminosity of 19.7 fb$^{-1}$ . 
Uncertainties include both statistical and systematic components, summed in quadrature.


%===================================================================
\begin{table}[h]
\centering
\label{ tab-marks }
\begin {tabular} {|l|c|}
\hline
process & No. of events \\
\hline
\hline
$DY \rightarrow \mu^{+} \mu^{-}$ & $533515~\pm~127708$ \\
\hline
$t\bar{t} +$ jets & $8363~\pm~2004$ \\
\hline
WW + jets & $1506~\pm~362.7$  \\
\hline
WZ + jets & $608~\pm~147.8$  \\
\hline
ZZ $\rightarrow 4\mu$ & $58~\pm~15.9$ \\
\hline
Sum Bkgs & $544050~\pm~130230$   \\
\hline
\hline
DH Signal & $8.2~\pm~3.5$ \\
(at $M_{Z^{\prime}}$ = 450 GeV) & \\ 
\hline
EFT Signal & $64.2~\pm~17.3$ \\
(at $\Lambda$ = 2 TeV) & \\ 
\hline
\end {tabular}
\caption{The number of dimuon events passing the preselection for each SM background processes, the DH model and the EFT model; corresponding to an integrated luminosity of 19.7 fb$^{-1}$. Uncertainties include both statistical and systematic components, summed in quadrature.}
\label{table:tab6}
\end{table}
%=====================================================================
\subsection{Events selection}
%\subsubsection{Final selection criteria}
After applying the preselection set of cuts, at which each event must have exactly two oppositely charged muons with $p^{\mu}_{T} >$ 45 GeV, $\eta^{\mu} <$ 2.1 each and one of these two muons should pass the single muon trigger (HLT$\_$Mu40$\_$eta2p1), the extra tighter selection has been optimized for DM signals in order to distinguish them from SM background and to obtain the best expected limit.
The final selection is based on three variables: 
(1) The mass of the dilepton system (M$_{\mu^{+}\mu^{-}}$) is required to be within $(0.9 \times M_{Z^{\prime}}) < M_{\mu^{+}\mu^{-}} < (M_{Z^{\prime}} + 25)$ to be consistent with leptons from the heavy Z$^{\prime}$ boson decay, 
(2) the azimuthal angle difference between the dimuon system and the missing transverse energy $\Delta\phi_{\mu^{+}\mu^{-},\vec{\slashed{p}}_{T}^{\text{corr}}}$ and 
(3) the relative difference between the dimuon system transverse momentum and the missing transverse momentum $|p_{T}^{\mu^{+}\mu^{-}} - \slashed{p}_{T}^{\text{corr}}|/p_{T}^{\mu^{+}\mu^{-}}$.
Here $p_{T}^{\mu^{+}\mu^{-}}$ is the dimuon transverse momentum, and $\Delta\phi_{\mu^{+}\mu^{-},\vec{\slashed{p}}_{T}^{\text{corr}}}$ is defined as difference in the azimuth angle between the dimuon system direction and the missing transverse momentum direction (i.e. $\Delta\phi_{\mu^{+}\mu^{-},\vec{\slashed{p}}_{T}^{\text{corr}}} = |\phi^{\mu^{+}\mu^{-}}-~\phi^{miss}|$ ) as indicated in table \ref{table:tabfinal}. \\

%At the final selection stage, in addition to the preliminary
%selection we open a mass window around the mas of the Z$^{\prime}$ and apply %extra cuts by making more constraints on the ratio $\Delta %P_{T}/P_{T}^{\mu\bar\mu}$, Where $\Delta P_{T}$ is the difference %between the dimuon transverse momentum ($P_{T}^{\mu\bar{\mu}}$) and %the missing transverse momentum ($P_{T}^{miss}$), and on the azimuth %angle between the dimuon and the missing transverse momentum as %indicated in table \ref{table:tabfinal}. \\
%\\

%==== Final selection ================
%\begin{table}[h!]
%    \centering
    %\begin{tabular}{|ll|}
%    \begin {tabular} {|l|c|}
%\hline
%variable & cut description \\
%\hline
%    \hline
%     Events passing the preliminary selection given in table %\ref{table:tabID1} & \\
%     Mass window at~: & $(0.9*M_{Z’}) < M_{\mu\bar\mu} < (M_{Z’} + %25)~Gev/c^{2}$ \\
%     $\Delta P_{T}/P_{T}^{\mu\bar{\mu}} < 0.6$~: & $\Delta P_{T} = %|P_{T}^{\mu\bar{\mu}} - P_{T}^{miss}|$  \\
%      $\Delta\phi > 2.6 $~: & $\Delta\phi = %|\phi^{\mu\bar{\mu}}-\phi^{miss}|$ \\
%    \hline
%    \end{tabular}
%    \caption{Summary of cut-based selection for the final analysis.}
%    \label{table:tabfinal}
%\end{table}

%==== Final selection ================ {zprime}
\begin{table*}[]
    \centering
    %\begin{tabular}{|ll|}
    \begin {tabular} {l c|c}
\hline
& variable & requirements \\
\hline
    \hline
    & Trigger     & HLT\_Mu40\_eta2p1 \\
    & High $p_{T}$ muon ID & \cite{R41, R32}\\
Preselection    & $p^{\mu}_{T}$ (GeV) & $>$ 45 \\
    & $\eta^{\mu}$ (rad) & $<$ 2.1 \\
    & $M_{\mu^{+}\mu^{-}}$ (GeV) & $>$ 50 \\
    
\hline
     &Mass window (GeV) & $(0.9 \times M_{Z^{\prime}}) < M_{\mu^{+}\mu^{-}} < (M_{Z^{\prime}} + 25)$ \\
Tight selection     &$|p_{T}^{\mu^{+}\mu^{-}} - \slashed{p}_{T}^{\text{corr}}|/p_{T}^{\mu^{+}\mu^{-}}$ & $<$ 0.6  \\
     &$\Delta\phi_{\mu^{+}\mu^{-},\vec{\slashed{p}}_{T}^{\text{corr}}}$ (rad) & $>$ 2.6 \\
    \hline
    \end{tabular}
    \caption{Summary of cut-based final events selection for the analysis.}
    \label{table:tabfinal}
\end{table*}

%======================================================================
\subsubsection{Selection efficiency} 
\begin{figure} [h]
\centering
\resizebox*{9cm}{!}{\includegraphics{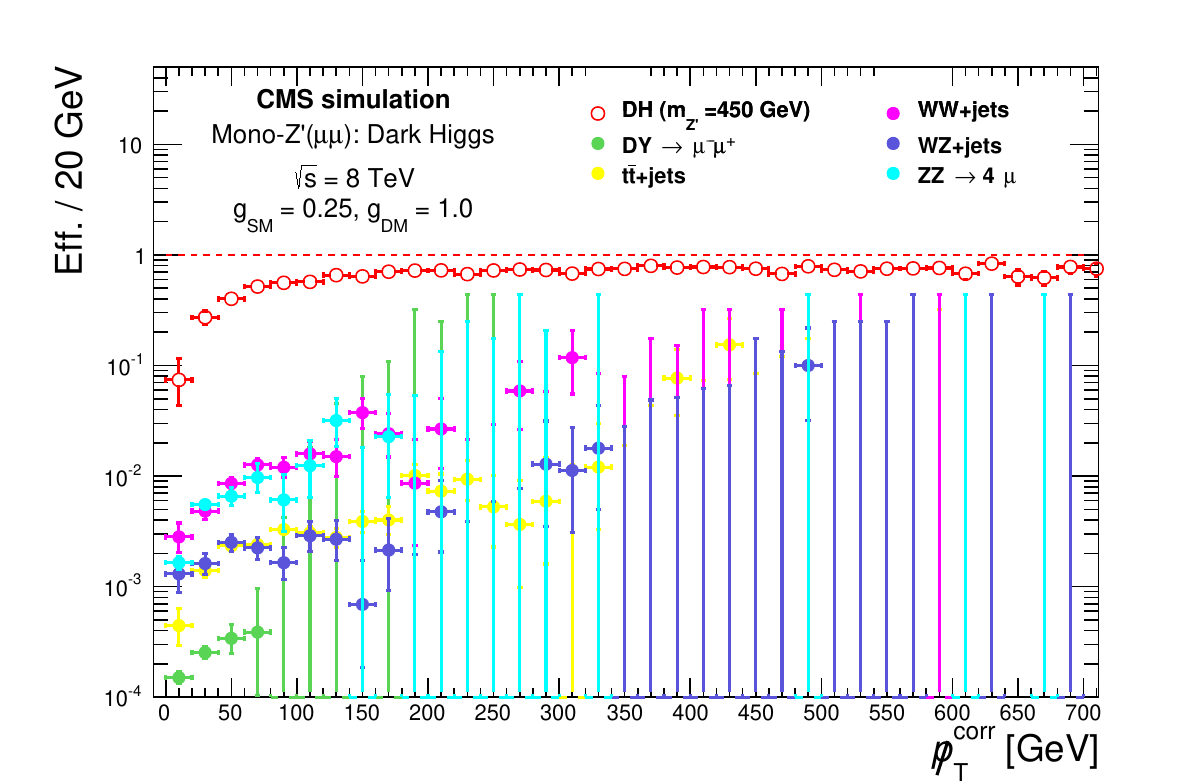}} %\resizebox*{10cm}{!}{\includegraphics{.png}}\hspac{}
  \caption{The efficiency presented as a function of $\slashed{\slashed{p}}_{T}^{~\text{corr}}$ 
  for the full selection of the analysis summarized in table \ref{table:tabfinal}; 
  for the SM backgrounds represented by solid dots with different corresponding colors and the DH model signal shown by hollow red dots.}
\label{figure:fig5}
\end{figure}

The selection efficiency is defined as the ratio between the number of events after applying the cut-based final events selection summarised in table \ref{table:tabfinal} and the number of events after the application of the preselection cuts, as defined in table \ref{table:tabID1}. 
These efficiencies are calculated for both dark Higgs scenario (with $M_{Z^{\prime}} = 450$ GeV) simulated sample and SM background sources, 
while error bars are statistical only.
The selection efficiencies are listed in percentage in table \ref{table:tab9}, and also shown in figure \ref{figure:fig5}.
The cut-based final events selection criteria described above is designed to reduce the background events with the minimal possible effect on the signal events. \\
After applying the full selection listed in table \ref{table:tabfinal}, most of the backgrounds events are strongly suppressed to less than a percent for each 
of the SM backgrounds, while we lose only around 34\% of the dark Higgs signal events. Nevertheless in the signal region ($\slashed{p}_{T}^{\text{corr}} > 200$ GeV) the efficiency of the DH signal is around 80\%, which proves the success of this selection criteria.
%=============================================================
%===========================================================

%%%%%% table + figure %%%%%%%%%%%%%%%%%%%%%%%%%%%%%
\begin{table}[h]
\centering
\label{ tab-marks }
\begin {tabular} {|c|c|}
\hline
\hline
Signal/Background & efficiency (\%)\\
\hline
DH signal & 67.0  \\
$DY \rightarrow \mu^{+} \mu^{-}$ & 0.0196 \\
$t\bar{t} +$ jets & 0.24 \\
WW + jets & 0.84 \\
WZ + jets & 0.20 \\
$ZZ \rightarrow 4\mu$  & 0.366\\
\hline
\hline
\end {tabular}
\caption{The over all efficiencies of the full selection, summarized in table \ref{table:tabfinal}, for the dark Higgs scenario signal calculated at $M_{Z'}=450$ GeV and the SM backgrounds.}
\label{table:tab9}
\end{table}
%======================================================================
\section{Systematic uncertainties}
\label{section:Uncertainties}
Several sources of experimental and theoretical systematic uncertainties are contributing to this analysis and affecting on the results. 
%However, the important point to be mentioned here is that the uncertainties have a %little impact on the driven limits as well as the exclusion ranges are set based on the %lack of events observed at those ranges.
We start with the experimental systematic uncertainties; these uncertainty related to the luminosity of the CMS-2012 data is estimated to be 2.6\% \cite{Lumi}. 
The uncertainty that arises from the determination of the muon detector acceptance and from the muon reconstruction efficiency ($A\times\epsilon$) has been found to be 3\% \cite{zprime}.
The transverse muon momentum resolution uncertainty was 5\%, while the misalignment in the detector geometry has an impact of 5\% on the transverse momentum scale uncertainty per TeV \cite{zprime}.
Regarding the systematic uncertainties on the measurements of missing transverse momentum $\slashed{\slashed{p}}_{T}^{~\text{corr}}$, the uncertainty in the energy scale of low energy particles which is known as unclustered energy was found to be 10\%, while 2-10\% for jet energy scale and 6-15\% for jet energy resolution \cite{R45}.
Finally the theoretical sources of the systematics are related to the uncertainties in the parton distribution functions (PDF) choice. 
For the Drell-Yan cross section calculation, this uncertainty can be represented 
as a function of the invariant mass of the dimuon as $(2.67+3.03\times10^{-3}M_{\mu^{+}\mu^{-}}+2.38\times10^{-6}M_{\mu^{+}\mu^{-}}^{2})\%$ (in GeV) \cite{zprime}. 
While PDF uncertainties for the WW and WZ processes were 5\% and 6\% respectively \cite{R450}.\\
The summary of these sources of uncertainties and the corresponding values are indicated in table \ref{table:sources}.

\begin{table}[h!]
    \centering
    \begin{tabular}{l r}
    \hline
    \hline
    Source     & Uncertainty (\%) \\
    \hline
    Luminosity ($\mathcal{L}$) & 2.6 {\footnotesize \cite{Lumi}} \\
    
    $A\times\epsilon$     & 3 {\footnotesize \cite{zprime}}\\
    
    $P_{T}$ resolution & 5 {\footnotesize \cite{zprime}}\\
    
    $P_{T}$ scale & 5 {\footnotesize \cite{zprime}}\\
   
    Unclustered $\slashed{\slashed{p}}_{T}^{~\text{corr}}$ scale & 10 {\footnotesize \cite{R45}} \\
    
    Jet energy scale & 2-10 {\footnotesize \cite{R45}} \\
    
    Jet energy resolution & 6-15 {\footnotesize \cite{R45}} \\
    
    PDF (Drell-Yan) & 4.5 {\footnotesize \cite{zprime}}\\
    
    PDF (ZZ) & 5 {\footnotesize \cite{R450}}\\
    
    PDF (WZ)  & 6 {\footnotesize \cite{R450}}\\ 
    \hline
    \hline
    \end{tabular}
    \caption{Different sources of systematic uncertainties estimated and the corresponding values.}
    \label{table:sources}
\end{table}

%=============================================================
%%%%\newpage textit
\section{Results}
\label{section:Results}
\begin{figure}[h]
\centering
  \resizebox*{9cm}{!}{\includegraphics{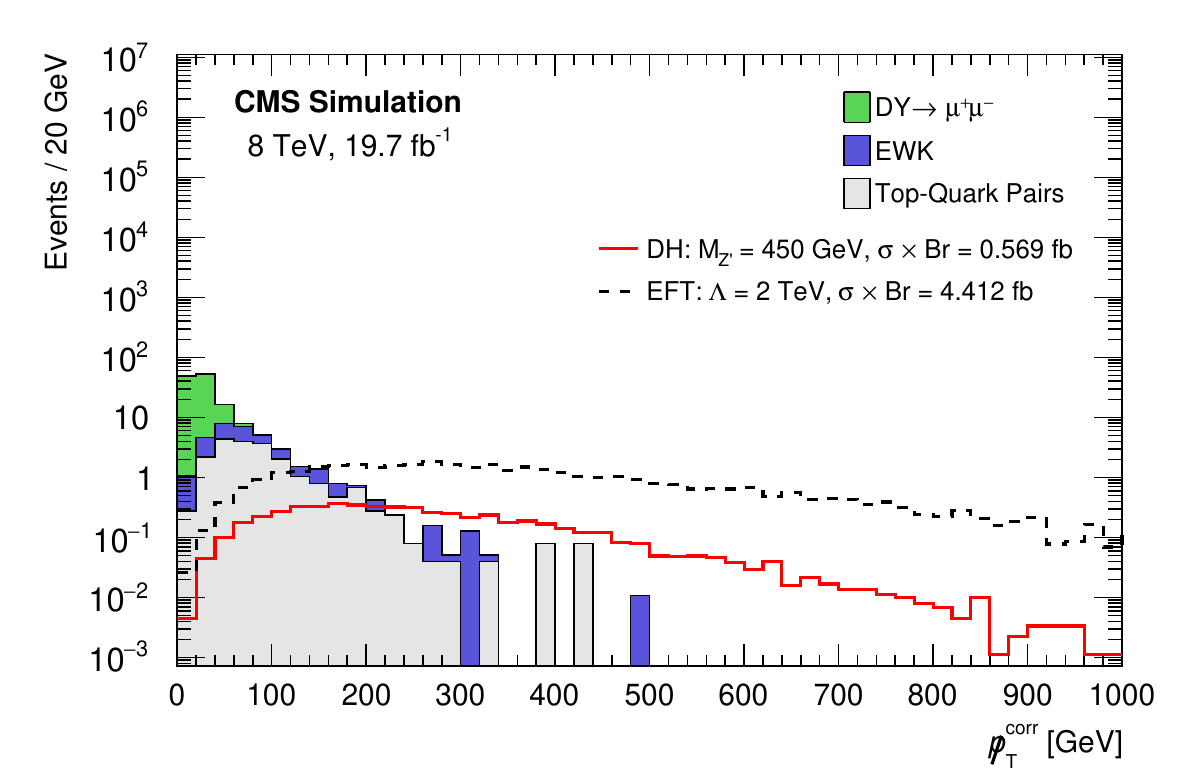}}
  \caption{The distribution of the missing transverse momentum, after the final selection cuts listed in table \ref{table:tabfinal}, for the SM background predictions, the signals of the model corresponding to the DH scenario with $M_{Z^{\prime}}=450$ GeV and the EFT scenario with cutoff scale $\Lambda=2$ TeV are superimposed. The signals are normalized to the product of cross section times the Z$^{\prime} \rightarrow \mu^{+}\mu^{-}$ branching ratio.}
%The systematic uncertainties, summarized in table \ref{table:sources}, are represented by the hatched band. The ratio between the data and the sum of the SM processes are illustrated in the bottom panel.}
  \label{figure:fig6}
\end{figure}

For the dimuon channel, a shape-based analysis is employed. 
The missing transverse momentum distributions ($\slashed{\slashed{p}}_{T}^{~\text{corr}}$) act as a good discriminant variable since the signals processes results in relatively larger  $\slashed{\slashed{p}}_{T}^{~\text{corr}}$ values than those of the SM backgrounds.\\
The missing transverse momentum distribution, after the final events selection, is shown in figure \ref{figure:fig6}, which shows a significant decrease of SM background processes with the use of the analysis final selection summarized in table \ref{table:tabfinal}.  
%The observed distribution agrees with the sum of the expected backgrounds within the statistical and systematic uncertainties and no excess of events is observed.\\
The number of dimuon events passing the final selection (summarized in \ref{table:tabfinal}) for each of the SM background processes, the DH model (with $M_{Z^{\prime}}$ = 450 GeV) and the EFT model (with $\Lambda=2~TeV$) corresponding to an integrated luminosity of 19.7 fb$^{-1}$ are shown in table \ref{table:tab8}. Uncertainties include both statistical and systematic components, summed in quadrature.
%%%%%%%%%%% plots step-2 %%%%%%%%%%%%%%%%%%%%%%%%%%%%%
%===================================================

%%%%%%%%%%%%%%% no of events %%%%%%%%%%%%%%%%%%%%%
\begin{table}[h!]
\centering
\label{ tab-marks }
\begin {tabular} {|l|c|}
\hline
Process & No. of events \\
\hline
\hline
$DY \rightarrow \mu^{+} \mu^{-}$ & $104.8~\pm~27.1$  \\
\hline
$t\bar{t} + jets$ & $20.3~\pm~6.6$\\
\hline
$WW + jets$ &$12.7~\pm~4.7$ \\
\hline
$WZ + jets$ &$1.2~\pm~1.2$ \\
\hline
$ZZ \rightarrow 4\mu$  & $0.2~\pm~0.5$   \\
\hline
Sum Bkgs & $139.1~\pm~35.3$  \\
\hline
\hline
Dark Higgs & $5.5~\pm~2.7$   \\
(at $M_{Z^{\prime}}$ = 450 GeV) &  \\ 
\hline
EFT & $39.2~\pm~11.3$   \\
(at $M_{Z^{\prime}}$ = 450 GeV) &  \\ 
\hline
\end {tabular}
\caption{The number of dimuon events passing the analysis final selection (summarized in table \ref{table:tabfinal}) for each SM backgrounds, the DH model and the EFT model; corresponding to an integrated luminosity of 19.7 fb$^{-1}$. Uncertainties include both statistical and systematic components, summed in quadrature.}
\label{table:tab8}
\end{table}

%--------------
%%%%\newpage
\subsection{Statistical interpretation}
In order to make a statistical interpretation of our results, we use the asymptotic approximation of the distribution of the profile likelihood-based statistical test described in details in \cite{R2}.
This approach was used to investigate the possibility of the rejection of the null hypothesis (the SM background only hypothesis) in favor of the signal hypothesis (the dark Higgs scenario hypothesis) and used to construct the confidence intervals within 1 or 2 standard deviation corresponding to 68\% or 95\% Confidence Levels (CL). 
The likelihood function used to fit the data is defined as  
% ++++++ Profile Likelihood equation +++++++
$$ \mathcal{L}(\mu,\theta) = \prod_{i=1}^{M} \frac{{(\mu s_{i} + b_{i})}^{n_{i}} ~e^{-(\mu s_{i} + b_{i})} }  {n_{i}~!}\prod_{j=1}^{k} \frac{ u_{j}^{m_{j}} ~e^{-u_{j}} }  {m_{j}~!}, $$
where $\mu$ is the signal strength, and defined as ration between the signal yield and those of the prediction from simulation, which is the parameter of interest (POI) in this analysis, and $\theta$ represents the other nuisance parameters with an impact included in the equation on the second $\Pi$-product.
While $s_i$ and $b_i$ are number of signal and background events, respectively, as estimated from MC simulation per each bin. 
Finally $u_{j}$ is a function of $\theta$ that gives the expectation value for each bin in the control sample used to constrain this nuisance parameters.
It is a shape analysis based on the missing transverse momentum distributions. 

\subsection{Exclusion limits}
\begin{figure} [H]
\centering
 \includegraphics[width=\linewidth]{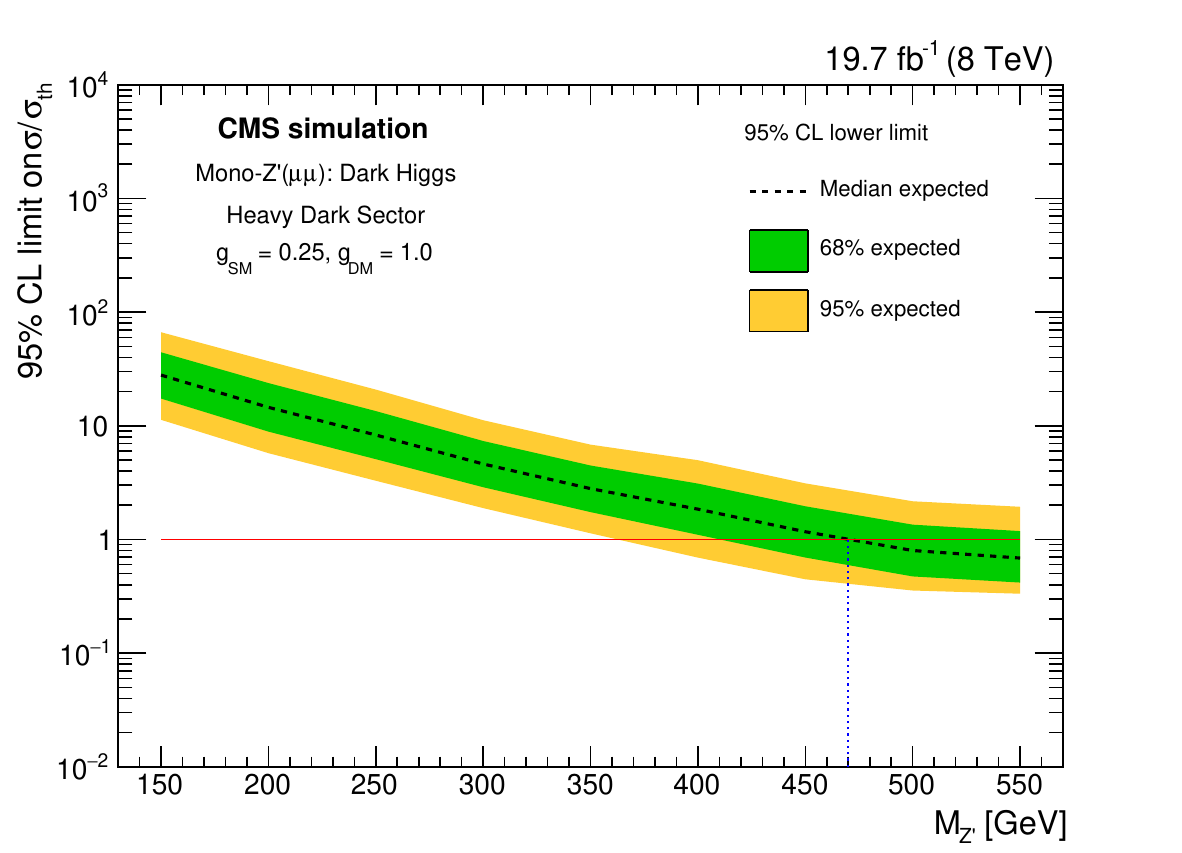}
  %\resizebox*{9cm}{!}{\includegraphics{fig2_DM_crosssection_vs_lambda_mc.pdf}}
  \caption{The limit at 95\% CL on the expected \text{$\sigma / \sigma_{\text{theory}}$} for the Dark Higgs scenario for the Z$^{\prime}$ dimuon decay of the Mono-Z$^{\prime}$ model. Distribution is shown as a function of $M_{Z'}$ for the heavy dark sector mass assumption shown in table 
  \ref{table:tab1}. The inner and outer shaded bands show the 68 and 95\% CL uncertainties in the expected limits, respectively. The horizontal red line refers to $\sigma / \sigma_{\text{theory}}$ = 1. The vertical blue dashed line point to the intersection of expectation with the case where $\sigma = \sigma_{\text{theory}}$.}
  \label{figure:limitMzp}
\end{figure}

\begin{figure} [H]
\centering
  \includegraphics[width=\linewidth]{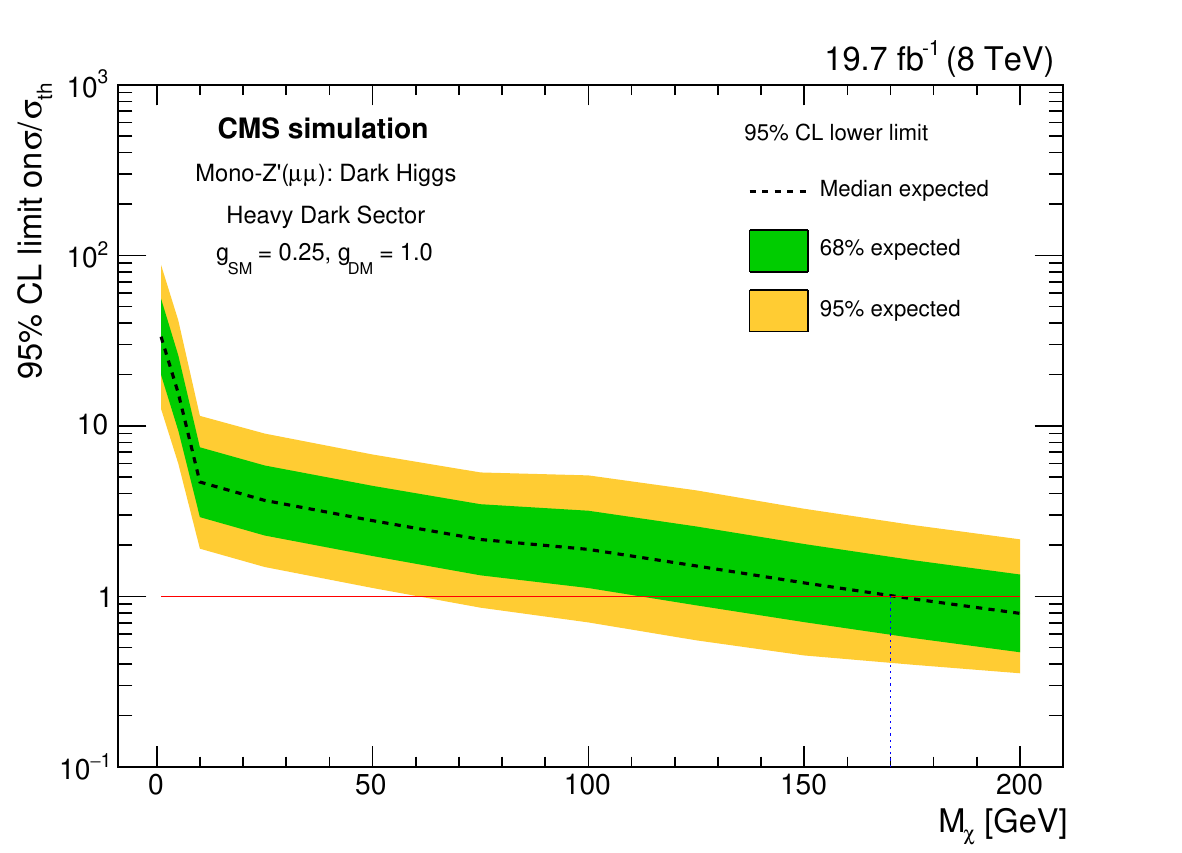}
  %\resizebox*{9cm}{!}{\includegraphics{limitMchi.pdf}}
  \caption{The limit at 95\% CL on the expected \text{$\sigma/\sigma_{\text{theory}}$} for the Dark Higgs scenario for the Z$^{\prime}$ dimuon decay of the Mono-${Z'}$ model. Distribution is shown as a function of $M_{\chi}$ for different values of the Z$^{\prime}$ mass. The inner and outer shaded bands show the 68 and 95\% CL uncertainties in the expected limits, respectively. 
  The horizontal red line refers to $\sigma / \sigma_{\text{theory}}$ = 1.
  The vertical blue dashed line point to the intersection of expectation with the case where $\sigma = \sigma_{\text{theory}}$.}
  \label{figure:limitMchi}
\end{figure}

\begin{figure} [H]
\centering
  \includegraphics[width=\linewidth]{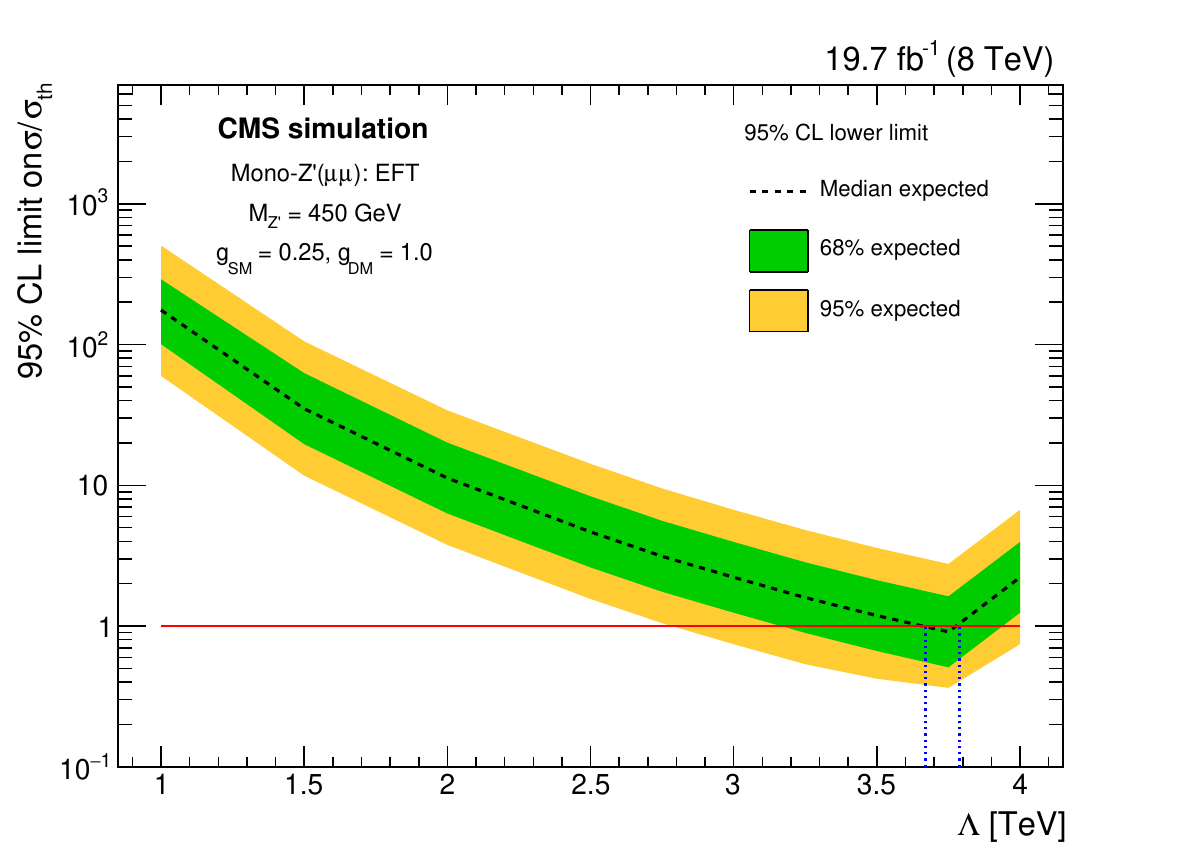}
  %\resizebox*{9cm}{!}{\includegraphics{limitMchi.pdf}}
  \caption{The limit at 95\% CL on the expected \text{$\sigma/\sigma_{\text{theory}}$} for the EFT scenario for the Z$^{\prime}$ dimuon decay of the Mono-${Z'}$ model. Distribution is shown as a function of the EFT cutoff scale ($\Lambda$). The inner and outer shaded bands show the 68 and 95\% CL uncertainties in the expected limits, respectively. 
  The horizontal red line refers to $\sigma / \sigma_{\text{theory}}$ = 1.
  The vertical blue dashed lines point to the intersection of expectation with the case where $\sigma = \sigma_{\text{theory}}$.}
  \label{figure:limitLambda}
\end{figure}

We construct the confidence intervals for the signal strengths as a function of the mass of the new $Z^{\prime}$ boson ($M_{Z^{\prime}}$) shown in figure \ref{figure:limitMzp}, and with the mass of the stable dark sector particle $M_{\chi}$ shown in figure \ref{figure:limitMchi} for the DH simplified scenario, the confidence intervals for the signal strengths as a function of $\Lambda$ for the EFT approach of the theory is presented in figure \ref{figure:limitLambda} as well.
%It is clearly shown that the simulation of the SM background is in a good agreement with the data within $\pm 2\sigma$, i.e. no deviation from the SM processes have been observed.
We exclude $Z^{\prime}$ production in the mass range between 470 - 550 GeV from the expected median as illustrated in figure \ref{figure:limitMzp}.
We also exclude DM particle ($\chi$) production in the mass range between 170 - 200 GeV from expected median as shown in figure \ref{figure:limitMchi}. For the EFT scenario, the range between 3670 - 3790 GeV is excluded for the model cutoff scale of the EFT($\Lambda$), as shown in figure \ref{figure:limitLambda}. These exclusion limits are estimated at 95\% confidence level.

% ========== exclusion limits plot =======================
%Observed (solid) and expected (dashed) 95\% CL lower limits on the %signal strength as a function of the mass of Z$^{\prime}$, within $\pm %1\sigma$ (green) and $\pm 2\sigma$ (yellow).
%=========================================
%================================================================
%=========================================
%================================================================
%%%\newpage
\section{Summary}
\label{section:Summary}
%A search for dark matter particles produced in association with a heavy gauge boson Z$^{\prime}$, using a sample of proton-proton collision CMS open data corresponding to an integrated luminosity of 19.7 fb$^{-1}$, has been performed. 

A search for dark matter particles produced in association with a heavy gauge boson Z$^{\prime}$ has been performed.
In this search we used MC samples which are taken from the open simulated files produced by the CMS collaboration for proton-proton collisions at $\sqrt{s} = $ 8 TeV, the analysis was optimized for the full LHC run-{\footnotesize I} integrated luminosity (19.7 fb$^{-1}$).
Results from muonic decay mode of Z$^{\prime}$ are presented, along with the statistical and systematic combination of uncertainties.
One benchmark signal corresponding to the dark Higgs scenario was used with different choices of the mediator (Z$^{\prime}$) mass points.
The 95\% CL limits on the expected \text{$\sigma/\sigma_{\text{theory}}$} of dark matter in a dark Higgs scenario extended by a Z$^{\prime}$ boson is set. 
These limits constitute the most stringent limits on the parameters ($M_{Z^{\prime}}$ and $M_{\chi}$) of this model.  
For the mass of ${Z'}$ ($M_{Z^{\prime}}$), the ranges between 470 - 550 GeV from expected median have been excluded, while excluding the ranges between 170 - 200 GeV from expected median for the mass of the dark matter $\chi$ ($M_{\chi}$). 
In addition, the effective field theory (EFT) formalism of the mono-Z' model was used for interpreting the result.  
Limit was set on EFT cutoff scale ($\Lambda$), thus $\Lambda$ range between 3.67 - 3.79 TeV is excluded.

We are planning to repeat the study with CMS full run-{\footnotesize II} data with lager centre-of-mass energy ($\sqrt{s}$ = 13 TeV) and data integrated luminosity ($\mathcal{L}$ = 137 fb$^{-1}$).\\
\\

\begin{acknowledgments}
M. A. Louka wish to acknowledge the support of the Centre for Theoretical Physics (CTP) at the British University in Egypt (BUE) 
for the financial support to this work. The authors of this paper would like to thank Tongyan Lin, one of the authors of \cite{R1}, for her useful discussions about the theoretical models, crosschecking of the results and for sharing with us the different scenarios Madgraph cards that were used for the events generation. 
We want to express our deepest thank to Nicola De Filippis from the politecnico di Bari/INFN for allowing us to use the computing facilities to produce and hosting our ntuples at Bari tier 2 servers. Finally we would like to thank Michele Gallinaro from LIP/Lisbon for his useful discussions about our results. 
\end{acknowledgments}

\nocite{*}

%========= References ===============================================


%apsrev4-2.bst 2019-01-14 (MD) hand-edited version of apsrev4-1.bst
%Control: key (0)
%Control: author (8) initials jnrlst
%Control: editor formatted (1) identically to author
%Control: production of article title (0) allowed
%Control: page (0) single
%Control: year (1) truncated
%Control: production of eprint (0) enabled
\begin{thebibliography}{0}%
\makeatletter
\providecommand \@ifxundefined [1]{%
 \@ifx{#1\undefined}
}%
\providecommand \@ifnum [1]{%
 \ifnum #1\expandafter \@firstoftwo
 \else \expandafter \@secondoftwo
 \fi
}%
\providecommand \@ifx [1]{%
 \ifx #1\expandafter \@firstoftwo
 \else \expandafter \@secondoftwo
 \fi
}%
\providecommand \natexlab [1]{#1}%
\providecommand \enquote  [1]{``#1''}%
\providecommand \bibnamefont  [1]{#1}%
\providecommand \bibfnamefont [1]{#1}%
\providecommand \citenamefont [1]{#1}%
\providecommand \href@noop [0]{\@secondoftwo}%
\providecommand \href [0]{\begingroup \@sanitize@url \@href}%
\providecommand \@href[1]{\@@startlink{#1}\@@href}%
\providecommand \@@href[1]{\endgroup#1\@@endlink}%
\providecommand \@sanitize@url [0]{\catcode `\\12\catcode `\$12\catcode
  `\&12\catcode `\#12\catcode `\^12\catcode `\_12\catcode `\%12\relax}%
\providecommand \@@startlink[1]{}%
\providecommand \@@endlink[0]{}%
\providecommand \url  [0]{\begingroup\@sanitize@url \@url }%
\providecommand \@url [1]{\endgroup\@href {#1}{\urlprefix }}%
\providecommand \urlprefix  [0]{URL }%
\providecommand \Eprint [0]{\href }%
\providecommand \doibase [0]{https://doi.org/}%
\providecommand \selectlanguage [0]{\@gobble}%
\providecommand \bibinfo  [0]{\@secondoftwo}%
\providecommand \bibfield  [0]{\@secondoftwo}%
\providecommand \translation [1]{[#1]}%
\providecommand \BibitemOpen [0]{}%
\providecommand \bibitemStop [0]{}%
\providecommand \bibitemNoStop [0]{.\EOS\space}%
\providecommand \EOS [0]{\spacefactor3000\relax}%
\providecommand \BibitemShut  [1]{\csname bibitem#1\endcsname}%
\let\auto@bib@innerbib\@empty
%</preamble>
\end{thebibliography}%


\begin{thebibliography}{9}

\bibitem{R4} F. Zwicky, \textit{The Redshift of the Extragalactic Nebulae}, Helv. Phys. Acta, Vol. 6, p. 110-127, 1933 \textcolor{blue}{[iNSPIRE-HEP]}.


\bibitem{R5} Yoshiaki SOFUE and Vera Rubin, \textit{Rotation Curves of Spiral Galaxies}, Ann.Rev.Astron.Astrophys. 39 (2001) 137-174 \textcolor{blue}{[arXiv:astro-ph/0010594] [iNSPIRE-HEP]}.

\bibitem{R6} Scherrer, Robert J. and Turner, Michael S.
\textit{On the relic, cosmic abundance of stable, weakly interacting massive particles}, 
Phys. Rev. D 33 (1986) 1585 \textcolor{blue}{[iNSPIRE-HEP]}.

\bibitem{R7} Planck Collaboration, \textit{Planck 2015 results. XIII. Cosmological parameters}, Astron. Astrophys. 594 (2016) A13  \textcolor{blue}{[arXiv:1502.01589] [iNSPIRE-HEP]}.


\bibitem{R8} Trimble, Virginia, \textit{Existence and Nature of Dark Matter in the Universe}, Annual Review of Astronomy and Astrophysics, Vol.25 (1987) 425-472 \textcolor{blue}{[iNSPIRE-HEP]}.

\bibitem{R9} Bertone, Gianfranco and Hooper, Dan and Si,
\textit{Particle dark matter:Evidence, candidates and constraints}, Phys. Rept. 405 (2005) 279-390 \textcolor{blue}{[arXiv:hep-ph/0404175] [iNSPIRE-HEP]}.

\bibitem{R10} L. Bergstrom,
\textit{Non-baryonic dark matter: observational evidence and detection methods}, Rept. Prog. Phys. 63 (2000) 793 \textcolor{blue}{[arXiv:hep-ph/0002126] [iNSPIRE-HEP]}.

\bibitem{R11} K. Abazajian, G. M. Fuller and M. Patel,
\textit{Sterile neutrino hot, warm, and cold dark matter}, Phys. Rev. D 64 (2001) 023501 \textcolor{blue}{[arXiv:astro-ph/0101524] [iNSPIRE-HEP]}.

\bibitem{R1010} Lage, C and Farrar, G, \textit{The bullet cluster is not a cosmological anomaly}, JCAP, vol. 2015, no. 2, 038. \textcolor{blue}{https://doi.org/10.1088/1475-7516/2015/02/038}.



%%%%%%%%%%%%%%

\bibitem{R35} CMS Collaboration, Search for new physics in final states with an energetic jet or a hadronically decaying W or Z boson and transverse momentum imbalance at $\sqrt{s}$ = 13~TeV, Phys. Rev. D 97 (2018) 092005. \textcolor{blue}{[arXiv:1712.02345] [hep-ex]}.

\bibitem {photon} CMS Collaboration, \textit{Search for new physics in the monophoton final state in proton-proton collisions at $\sqrt{s}$ = 13~TeV, JHEP. 10 (2017) 073}, \textcolor{blue}{[arXiv:1706.03794v2] [hep-ex]}. 

\bibitem{R36} CMS Collaboration,
\textit{Search for dark matter particles produced in association with a Higgs boson in proton-proton collisions at $\sqrt{s}$~=~13 TeV, JHEP 03 (2020) 025}, \textcolor{blue}{[arXiv:1908.01713v2] [hep-ex]}.

\bibitem{R450} CMS Collaboration,
\textit{Search for dark matter and unparticles produced in association with a Z boson in proton-proton collisions at $\sqrt{s}$ = 8 TeV}, Phys. Rev. D 93, 052011 (2016) \textcolor{blue}{[arXiv:1511.09375] [hep-ex]}.

\bibitem{R45055} CMS Collaboration,
\textit{Search for dark matter produced in association with a leptonically decaying Z boson in proton-proton collisions at $\sqrt{s}$ = 13 TeV}, Eur. Phys. J. C 81 (2021) 13 \textcolor{blue}{[arXiv:2008.04735] [hep-ex]}.

%%%%%%%%%%%%%%%%%%%%
\bibitem{R12} Boveia, Antonio and Doglioni, Caterina,
\textit{Dark Matter Searches at Colliders}, Ann. Rev. Nucl. Part. Sci. 68 (2018) 429-459 \textcolor{blue}{[arXiv:1810.12238] [hep-ex]}.

\bibitem{R38} Krovi Anirudh, Low Ian and Zhang Yue,
dark matter searches at the LHC: mono-X versus darkonium channels. JHEP 10 (2018) 026 \textcolor{blue}{[arXiv:1807.07972] [hep-ph]}.

\bibitem{R1}
 Marcelo Autran, Kevin Bauer, Tongyan Lin and Daniel Whiteson,
\textit{Mono-Z$^{\prime}$: searches for dark matter in events with a resonance and missing transverse energy}. Physical Review D 92 (2015) 035007 \textcolor{blue}{[arXiv:1504.01386] [hep-ph]}.

%%%%%%%%%
\bibitem{R13} Paul Langacker,
\textit{The physics of heavy $Z^{\prime}$ gauge bosons}, Rev. Mod. Phys. 81 (2009) 1199-1228 \textcolor{blue}{[arXiv:0801.1345] [hep-ph]}.

\bibitem{R14} Cheng-Wei Chiang, Takaaki Nomurad and Kei Yagyu,
\textit{Phenomenology of E 6-inspired leptophobic $Z^{\prime}$ boson at the LHC}, JHEP (2014) \textcolor{blue}{[arXiv:1402.5579] [hep-ph]}.

\bibitem{R15} Digesh Raut,
\textit{gauged U(1) extension of the SM and Phenomenology}, PhD U. Alabama, Tuscaloosa (2018) \textcolor{blue}{[iNSPIRE-HEP]}.

\bibitem{R16} Robert Foot, X.G. He , H. Lew and R.R. Volkas,
\textit{Model for a light Z-prime boson}, Phys.Rev.D 50 (1994) 4571-4580
\textcolor{blue}{[arXiv:9401250][hep-ph]}.

\bibitem{R37} ATLAS Collaboration,
\textit{Search for dark matter in events with a hadronically decaying vector boson and missing transverse momentum in pp collisions at $\sqrt{s}$ = 13~TeV with the ATLAS detector}, JHEP 10 (2018) 180 \textcolor{blue}{[arXiv:1807.11471] [hep-ex]}.

\bibitem {LEP} J. Alcaraz et al. (ALEPH Collaboration, DELPHI Col-
laboration, L3 Collaboration, OPAL Collaboration, LEP
Electroweak Working Group) (2006), hep-ex/0612034.

\bibitem{R21} The CMS Collaboration, Software Framework for CMS Open Data Analysis, \textcolor{blue}{http://opendata.cern.ch/docs/about-cms.}

\bibitem{R3} Aram Apyan, William Cuozzo,  Markus Klute, Yoshihiro Saito, Matthias Schott and Bereket Sintayehu,
\textit{Opportunities and challenges of Standard Model production cross section measurements in proton-proton collisions at $\sqrt{s}=8$~TeV using CMS Open Data}, JINST 15 (2020) \textcolor{blue}{[arXiv:1907.08197] [hep-ex]}.

\bibitem{R33} J. Alwall, R. Frederix, S. Frixione, V. Hirschi, F. Maltoni, O. Mattelaer, H.-S. Shao, T. Stelzer, P. Torrielli \& M. Zaro,
\textit{The automated computation of tree-level and next-to-leading order differential cross sections, and their matching to parton shower simulations}. JHEP 07 (2014) 079 \textcolor{blue}{[arXiv:1405.0301] [hep-ph]}.

\bibitem{R17} CMS collaboration,
\textit{The CMS Experiment at the CERN LHC}, JINST 3 (2008) S08004 \textcolor{blue}{[iNSPIRE-HEP]}.

\bibitem{R29} CMS Physics, Technical Design Report Volume I: Detector Performance and Software, 2006.

%\bibitem{R18} M.Mulders, CMS collaboration,
%\text{Muon Reconstruction and Identification at CMS}, Nuclear Physics B 172 %(2007) 205-207.

\bibitem{R18}
M. Mulders,
\textit{Muon Reconstruction and Identification at CMS},
Nuclear Physics B - Proceedings Supplements,
Volume 172,
2007,
Pages 205-207,
ISSN 0920-5632,
\textcolor{blue}{https://doi.org/10.1016/j.nuclphysbps.2007.08.049}.

\bibitem{R40} CMS Collaboration,
\textit{Performance of CMS muon reconstruction in pp collision events at $\sqrt{s}$ = 7~TeV}, JINST 7 (2012) \textcolor{blue}{[arXiv:1206.4071] [physics.ins-det]}.
%=================================================================================
\bibitem{R19} CMS collaboration,
Particle-Flow Event Reconstruction in CMS and Performance for Jets, Taus, and MET, Tech. Rep. CMS-PAS-PFT-09-001, CERN, Geneva, Apr, 2009.
%================================================================================

\bibitem{R45} CMS Collaboration,
\textit{Performance of the CMS missing transverse energy reconstruction in pp data at $\sqrt{s}$ = 8~TeV, JINST 10 (2015) P02006}, \textcolor{blue}{[arXiv:1411.0511] [physics.ins-det]}.

\bibitem{R34}  Torbjon sjostrand, stephen Mrenna and peter skands,
\textit{PYTHIA 6.4 Physics and Manual}, JHEP 05 (2006) 026 \textcolor{blue}{[arXiv:hep-ph/0603175]}.

\bibitem{powheg} E. Re, Single-top Wt-channel production matched with parton showers using the POWHEG method, Eur. Phys. J. C 71 (2011) 1547 \textcolor{blue}{[arXiv:1009.2450] [INSPIRE]}.

\bibitem{Ropendata} CMS Collaboration, Thomas McCauley,
\textit{Open Data at CMS: Status and Plans.}, PoS LHCP2019 (2019) 260 \textcolor{blue}{[iNSPIRE-HEP]}.

\bibitem{zprime} CMS Collaboration,
\textit{Search for physics beyond the standard model in dilepton mass spectra in proton-proton collisions at $\sqrt{s}$ = 8~TeV}, JHEP 04 (2015) 025 \textcolor{blue}{[arXiv:1412.6302] [hep-ex]}.

\bibitem{R39} CMS Collaboration,
\textit{CMS list of validated runs for primary datasets of 2012 data taking},
CERN Open Data Portal. \textcolor{blue}{DOI:10.7483/OPENDATA.CMS.C00V.SE32},

%============= CMS MCs Samples ==============
\bibitem{R22} CMS collaboration, 
Simulated dataset DYToMuMu\_M-20\_CT10\_8TeV-powheg-pythia6 in AODSIM format for 2012 collision data. CERN Open Data Portal: \textcolor{blue}{http://opendata.cern.ch/record/774},


\bibitem{R23} CMS collaboration, 
Simulated dataset TTJets\_FullLeptMGDecays\_8TeV-madgraph in AODSIM format for 2012 collision data. CERN Open Data Portal: \textcolor{blue}{http://opendata.cern.ch/record/9577}.

\bibitem{R24} CMS collaboration, 
Simulated dataset WWJetsTo2L2Nu\_TuneZ2star\_8TeV-madgraph-tauola in AODSIM format for 2012 collision data. CERN Open Data Portal: \textcolor{blue}{http://opendata.cern.ch/record/9971}.


\bibitem{R25} CMS collaboration, 
Simulated dataset WZJetsTo3LNu\_TuneZ2\_8TeV-madgraph-tauola in AODSIM format for 2012 collision data. CERN Open Data Portal: \textcolor{blue}{http://opendata.cern.ch/record/9983}.


\bibitem{R26} CMS collaboration, 
Simulated dataset ZZTo4mu\_8TeV-powheg-pythia6 in AODSIM format for 2012 collision data. CERN Open Data Portal:
\textcolor{blue}{http://opendata.cern.ch/record/10071}.

%%%%%%%%%%%%%%%%
%%%%% CMS Data %%%%%%%%%
%\bibitem{R27} CMS collaboration, 
%SingleMu primary dataset in AOD format from Run of 2012 %(/SingleMu/Run2012B-22Jan2013-v1/AOD). CERN Open Data Portal:
%\textcolor{blue}{http://opendata.cern.ch/record/6021}.

%\bibitem{R28} CMS collaboration, 
%SingleMu primary dataset in AOD format from Run of 2012 %(/SingleMu/Run2012C-22Jan2013-v1/AOD). CERN Open Data Portal:
%\textcolor{blue}{http://opendata.cern.ch/record/6047}.
%%%%%%%%  ID
\bibitem{R41} CMS Collaboration,
\textit{Search for Resonances in the Dilepton Mass Distribution in pp Collisions at $\sqrt{s}$ = 7~TeV}, JHEP 1105 093 (2011) \textcolor{blue}{[arXiv:1103.0981] [hep-ex]}.

\bibitem{R32}
{https://twiki.cern.ch/twiki/bin/view/CMSPublic/SWG
uideMuonId\#HighPT\_Muon}.
%%%%%%

\bibitem{R31} Alexender Spiridonov,
\textit{An Approch To Global Vertex Fitting}, DESY-IfH Zeuthen / IHEP Protvino.

%\bibitem{R30} R. Frohwirth, 
%\text{Application Of Kalman Filtering to Track And Vertex Fitting}, %Nucl.Instrum.Meth.A 262 (1987).

\bibitem{R30} R. Frühwirth,
\textit{Application of Kalman filtering to track and vertex fitting},
Nuclear Instruments and Methods in Physics Research Section A: Accelerators, Spectrometers, Detectors and Associated Equipment,
Volume 262, Issues 2–3,
1987,
Pages 444-450,
ISSN 0168-9002,
\textcolor{blue}{https://doi.org/10.1016/0168-9002(87)90887-4}.
%%%%%%%%

\bibitem{Lumi}
CMS Collaboration, \textit{CMS Luminosity Based on Pixel Cluster Counting - Summer 2013 Update}, CMS Physics Analysis Summary CMS-PAS-LUM-13-001 (2013).

%\bibitem{R450} CMS Collaboration,
%Search for dark matter and unparticles produced in association with a Z boson in %proton-proton collisions at $\sqrt{s}$ = 8~TeV, Phys. Rev. D 93, 052011 (2016) %\textcolor{blue}{[arXiv:1511.09375] [hep-ex]}.

\bibitem{R2} Glen Cowan , Kyle Cranmer , Eilam Gross , Ofer Vitells,
\textit{Asymptotic formulae for likelihood-based tests of new physics}, Eur. Phys. J. C 71, 1554 (2011) \textcolor{blue}{[arXiv:1007.1727] [physics.data-an]}.



\end{thebibliography}
\end{document}